\documentclass[a4paper,11pt]{article}

\usepackage{jheppub} 
\usepackage[vcentermath]{youngtab}
\usepackage[T1]{fontenc} 

\usepackage{xcolor}
\usepackage{pstcol,pst-node}
\usepackage{graphics,epstopdf}
\usepackage{epic}
\usepackage{curves}
\usepackage[matrix,arrow,frame,curve]{xy}
\CompileMatrices
\usepackage{bbm}

\newcommand{\CF}{\mathcal{F}}

\newcommand{\CN}{\mathcal{N}}
\newcommand{\CO}{\mathcal{O}}

\newcommand{\CS}{\mathcal{S}}

\newcommand {\apgt} {\ {\raise-.5ex\hbox{$\buildrel>\over\sim$}}\ }
\newcommand {\aplt} {\ {\raise-.5ex\hbox{$\buildrel<\over\sim$}}\ }

\makeatletter\@addtoreset{equation}{section}\makeatother

\newcommand{\Tr}{{\rm Tr\,}}

\def\Tr{\mathop{\rm Tr}}

\def\Z{\relax\ifmmode\mathchoice
{\hbox{\cmss Z\kern-.4em Z}}{\hbox{\cmss Z\kern-.4em Z}} {\lower.9pt\hbox{\cmsss Z\kern-.4em Z}}
{\lower1.2pt\hbox{\cmsss Z\kern-.4em Z}}\else{\cmss Z\kern-.4em Z}\fi}

\def\btimes{~{{{\lower1pt\hbox{$\square$}} \kern-7.6pt \times}}~}

\def\Z{{\Bbb{Z}}}

\def\be{\begin{equation}}
\def\ee{\end{equation}}
\newcommand{\bea}{\begin{eqnarray}}
\newcommand{\eea}{\end{eqnarray}}

\def\Tr{{\rm Tr}}

\renewcommand{\bar}{\overline}
\renewcommand{\hat}{\widehat}
\renewcommand{\tilde}{\widetilde}

\makeatletter
\newcommand{\ellSN}{\mathop{\operator@font sn}\nolimits}
\newcommand{\ellCN}{\mathop{\operator@font cn}\nolimits}
\newcommand{\ellDN}{\mathop{\operator@font dn}\nolimits}
\newcommand{\ellAM}{\mathop{\operator@font am}\nolimits}
\newcommand{\ellK}{\mathop{\smash{\operator@font K}\vphantom{a}}\nolimits}
\newcommand{\ellE}{\mathop{\smash{\operator@font E}\vphantom{a}}\nolimits}
\makeatother

\ifx\genfrac\sdflkaj

\else

\fi

\newcommand{\beq}{\begin{equation}}
\newcommand{\eeq}{\end{equation}}

\makeatletter
\def\mr@ignsp#1 {\ifx\:#1\@empty\else #1\expandafter\mr@ignsp\fi}%
\newcommand{\multiref}[1]{\begingroup
\xdef\mr@no@sparg{\expandafter\mr@ignsp#1 \: }%
\def\mr@comma{}%
\@for\mr@refs:=\mr@no@sparg\do{\mr@comma\def\mr@comma{,}\ref{\mr@refs}}%
\endgroup}
\makeatother

\newcommand{\hypref}[2]{\ifx\href\asklfhas #2\else\href{#1}{#2}\fi}

\renewcommand{\eqref}[1]{(\multiref{#1})}

\def\[{\begin{equation}}
\def\]{\end{equation}}
\def\<{\begin{eqnarray}}
\def\>{\end{eqnarray}}

\title{On Painlev\'{e}/gauge theory correspondence}

\author[a]{Giulio Bonelli,}
\author[b]{Oleg Lisovyy,}
\author[c]{Kazunobu Maruyoshi,}
\author[d]{Antonio Sciarappa,}
\author[a]{and Alessandro Tanzini}

\affiliation[a]{International School of Advanced Studies (SISSA)\\ 265 via Bonomea, Trieste, 34136, Italy}
\affiliation[b]{Laboratoire de Math\'ematiques et Physique Th\'eorique CNRS/UMR 7350 \\
Universit\'e de Tours, Parc de Grandmont, 37200 Tours, France}
\affiliation[c]{Faculty of Science and Technology, Seikei University \\
3-3-1 Kichijoji-Kitamachi, Musashino-shi, Tokyo, 180-8633, Japan}
\affiliation[d]{School of Physics, Korea Institute for Advanced Study \\
85 Hoegiro, Dongdaemun-gu, Seoul 130-722, Republic of Korea}

\emailAdd{bonelli@sissa.it}
\emailAdd{lisovyi@lmpt.univ-tours.fr}
\emailAdd{maruyoshi@st.seikei.ac.jp}
\emailAdd{asciara@kias.re.kr}
\emailAdd{tanzini@sissa.it}

\abstract{We elucidate the relation between Painlev\'{e} equations and four-dimensional rank one $\mathcal{N} = 2$ theories by identifying the connection associated to Painlev\'{e} isomonodromic problems with the oper limit of the flat connection of the Hitchin system associated to gauge theories and by studying the corresponding renormalisation group flow. 
Based on this correspondence we provide long-distance expansions at various canonical rays for all Painlev\'{e} $\tau$-functions in terms of magnetic and dyonic Nekrasov partition functions for $\mathcal{N} = 2$ SQCD and Argyres-Douglas theories at self-dual Omega background $\epsilon_1 + \epsilon_2 = 0$, or equivalently in terms of $c=1$ irregular conformal blocks.}

\preprint{KIAS-P16087}

\begin{document}

\maketitle

\section{Introduction} 
\label{Intro}

The non-perturbative formulation of gauge theories naturally links to rich mathematical structures underlying classical and quantum integrable systems, representation theory of infinite dimensional algebrae and moduli spaces. This is particularly evident in the context of $\mathcal{N}=2$ supersymmetric gauge theories in four dimensions, where a plethora of exciting results have been obtained since the seminal work of Seiberg and Witten \cite{Seiberg:1994rs,Seiberg:1994aj}.
A particularly interesting class of such theories, dubbed class $\mathcal{S}$ \cite{Gaiotto:2009we, 2009arXiv0907.3987G}, is the one that can be obtained by compactifying M-theory on $\mathbb{R}^7 \times Q$, where $Q$ is a holomorphic symplectic manifold, with $N$ M5-branes on $\mathbb{R}^4\times \mathcal{C}$, $\mathcal{C}$ being a holomorphic two-cycle in $Q$ \cite{Witten:1997sc}. Since we are interested only in the low-energy effective M-theory neglecting the gravity dynamics, we focus on a tubular neighbourhood of the two-cycle, whose local geometry is described by the total space of the cotangent bundle $T^*\mathcal{C}$. In this setting, the Coulomb branch of the resulting four-dimensional 
theory is parametrised by the commuting Hamiltonians of an associated Hitchin's integrable system, which are the moduli of meromorphic differentials on the curve $\mathcal{C}$ \cite{Donagi:1995cf}. 

In this paper, we will consider superconformal and asymptotically free $SU(2)$ gauge theories, and their Argyres-Douglas points \cite{Argyres:1995jj, Argyres:1995xn} arising from the compactification of two M5 branes, which are described in terms of quadratic differentials with respectively regular and irregular singularities.   
We will show that the renormalization group equations for these theories are described by Painlev\'{e} equations.
We focus on the strongly coupled phases of the four-dimensional theories which correspond to the long-distance regime of Painlev\'{e} solutions. This correspondence is shown from the following two complementary viewpoints. 

The first consists in identifying the punctured curve $\mathcal{C}$ with the auxiliary space where the isomonodromic problem associated to the Painlev\'{e} equation is formulated.
Consequently the Hitchin field gets identified with the connection of the spectral problem and the classification of the corresponding Painlev\'{e} equations mirrors that of the four-dimensional theories. 
Indeed, the coalescence diagram of Painlev\'{e} equations is identical to the one obtained in the gauge theories by puncture collisions. The isomonodromic deformations correspond to the Whitham deformations of the Hitchin system
\cite{Levin:1997tb,Teschner:2010je} which are in turn identified in the four-dimensional theory with the renormalization group equations \cite{Gorsky:1995zq, Edelstein:1999xk}. 

The second is the identification of the Nekrasov--Okounkov dual partition function of the gauge theory with the corresponding Painlev\'{e} $\tau$-function.
A particularly interesting class of theories that we consider are the Argyres-Douglas (AD) ones which are recognised to be in one-to-one correspondence with specific Painlev\'{e} equations. The analysis of these cases leads to new long-distance expansions of the Painlev\'{e} $\tau$-functions. This construction, via AGT correspondence, links the latter ones to irregular conformal blocks of Virasoro algebra \cite{Bonelli:2011aa,Felinska:2011tn,Gaiotto:2012sf,Kanno:2012xt,Nishinaka:2012kn}.

The paper is organized as follows. 
In section \ref{sw} we describe the $M$-theoretic origin of the Painlev\'{e}/gauge theory correspondence and identify the 
Seiberg-Witten curves with the spectral curves of the Painlev\'{e} systems.
In section \ref{Painleve} we describe new long-distance expansions of the Painlev\'{e} $\tau$-functions, focusing on the ones related to the AD  theories in four dimensions.
In section \ref{gauge} we compute the dual prepotentials of the AD theories in Coulomb branch in the self-dual $\Omega$-background.
Section \ref{concl} contains some concluding remarks and open questions.
We collect in the appendices the explicit computations for the Lagrangian gauge theories in the dyonic phases and the corresponding long-distance Painlev\'{e} expansions.

\section{Seiberg-Witten curves and isomonodromic deformations}
\label{sw}

In this Section we establish a correspondence between Seiberg-Witten (SW) geometry 
of four-dimensional $\mathcal{N} = 2$ class $\mathcal{S}$ theories constructed from two M5-branes and the system of linear Ordinary Differential Equations (ODE) associated to Painlev\'{e} equations, along the following lines:
\begin{itemize}
\item
we recall the linear ODE problems associated to the Painlev\'{e} equations, the Lax pair in suitable coordinates
and the corresponding Hamiltonian functions;
\item
we recall basic facts about Hitchin systems and their link to 4d $\mathcal{N} = 2$ gauge theories;
\item
we identify Hitchin's connection with the connection of the linear system and consequently the SW quadratic differential with the generator of Painlev\'{e} Hamiltonians;
\item
we recall the relation between Painlev\'{e} $\tau$-functions, conformal blocks and ``dual'' instanton partition functions.
\end{itemize}

\subsection{Painlev\'{e} equations and isomonodromic deformations}
At the turn of the twentieth century there have been repeated attempts to describe ordinary differential equations in terms of singularities of their solutions in the complex domain. For nonlinear ODEs this task is quite difficult to tackle, as in general the positions of singularities and even their type depend on the initial conditions: the singular points become \textit{movable}. A more refined version of the problem is to classify the equations with predictable branching by allowing only movable poles, movable single-valued essential singularities and fixed singular points of any type. The latter requirement is nowadays known as the \textit{Painlev\'{e} property}.

The classification of algebraic 1st order ODEs of Painlev\'{e} type has been achieved by L. Fuchs in 1884, who showed that such equations are either reducible to linear ones or can be solved in terms of elliptic functions. The treatment of the problem is simplified by the fact that movable singular points of the 1st order equations can only be poles or algebraic branch points. When the order of an ODE is 2 or higher, one cannot a priori exclude more complicated movable singularities, such as essential singular points or natural boundaries, and the problem becomes much more involved. Nevertheless in 1900-1910 P.~Painlev\'{e} and B. Gambier undertook an attempt of classifying the degree 1 2nd order ODEs free of movable branch points. Such equations have the form
\begin{equation}
\ddot{q} = F(q,\dot{q};t),
\end{equation}
where dots denote derivatives with respect to $t$ and $F$ is rational in $q$, $\dot q$ and locally analytic in $t$. The outcome of Painlev\'{e} and Gambier studies was a list of 50 equivalence classes of ODEs, most of which can be reduced to linear equations or integrated by quadratures. There remain six irreducible exceptional ODEs which became known under the name of Painlev\'{e} equations PI--PVI\footnote{See for example \cite{Fokas:2006:PTR} for an introduction to the subject.}.

\begin{figure}[h]
\centering
\includegraphics[height=3.5cm]{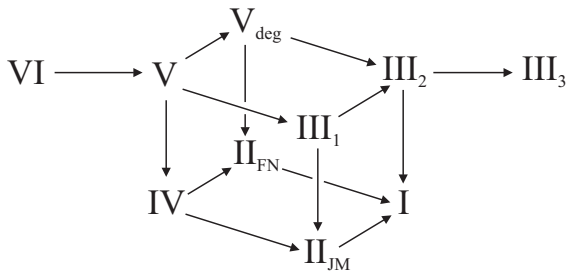}
\caption{Coalescence diagram for Painlev\'{e} equations.} \label{coalescence}
\end{figure}

Later studies based on the analysis of symmetries \cite{Okamoto1986I,198747,Okamoto1986,Ok4,OKSO} and spaces of initial conditions \cite{19791,Sakai2001} led to a refinement of the original classification into ten equations PI, PII$_{JM}$, PII$_{FN}$, PIII$_3$, PIII$_2$, PIII$_1$, PIV, PV$_{deg}$, PV and PVI. In this refinement PIII$_{1,2,3}$ correspond to specializations of the general Painlev\'{e} III equation
\footnote{A more conventional labeling of different PIII equations uses the rational surfaces desribing appropriate spaces of initial conditions; in this notation, ${\rm PIII}_1={\rm PIII}(D_6)$, ${\rm PIII}_2={\rm PIII}(D_7)$, ${\rm PIII}_3={\rm PIII}(D_8)$.
 } while PV$_{deg}$ and PII$_{FN}$ can be mapped respectively to PIII$_1$ and PII$_{JM} = $ PII upon change of variables (although the associated Lax pairs are different), and for this reason we will often not distinguish them. All these equations are related by \textit{coalescence}, that is they can be obtained from PVI by sequences of scaling limits according to the diagram depicted in Figure \ref{coalescence}. As it will be important for our following discussion, let us also mention here that Painlev\'{e} equations depend on a finite number of free parameters as listed in Table \ref{numpar}; the explicit expressions for all the equations will be given in the next sections and appendices. \\ 

\begin{center}
\begin{table}[h]
\centering
\begin{tabular}{|c||c|c|c|c|c||c|c|c|c}\hline
  & PVI & PV & PIII$_1$& PIII$_2$ & PIII$_3$ & PIV & PII & PI \\ 
\hline\hline $\text{n}^{\circ}$ parameters & 4 & 3 & 2 & 1 & 0 & 2 & 1 & 0 \\ \hline
\end{tabular}
\caption{Number of free parameters in the various Painlev\'{e} equations.} \label{numpar}
\end{table}
\end{center} 

There are two important realizations of the Painlev\'{e} equations that will help us understand their connection to supersymmetric gauge theories. The first realization is in terms of non-autonomous Hamiltonian systems, that is Painlev\'{e} equations arise as the equations of motion of a classical mechanics system with a \textit{time-dependent} Hamiltonian \cite{zbMATH02598485,Okamoto1999}
\begin{equation}
 \dot q 
     =     \frac{\partial}{\partial p} H_J(q,p;t), ~~~
    \dot p 
     =   - \frac{\partial}{\partial q} H_J(q,p;t)
          . 
\end{equation}
where ${\rm J~=~I, II}, {\rm III}_i, {\rm IV, V, VI} ~(i=1,2,3)$; the relevant Hamiltonians can be found in \cite{Okamoto1999,OKSO}.
Because of their time dependence, we can study the time evolution of our Hamiltonians; in the literature this is done by introducing the $\sigma_J(t)$ functions defined as
\begin{equation}
\sigma_J(t) = t\, H_J(t) ~~~ (J = {\rm III}_i, {\rm V}), ~~~~~~~~~
\sigma_J(t) = H_J(t) ~~~ (J = {\rm I}, {\rm II}, {\rm IV}). 
\end{equation}
together with a similar but slightly more complicated relation for PVI. These functions satisfy the $\sigma$-form of Painlev\'{e} equations, which also enter in the classification of degree two second order ordinary linear differential equations with Painlev\'{e} property \cite{Cosgrove}; we will write down these equations case by case in the following sections of this paper. It is also useful to consider the so-called $\tau$-functions $\tau_J(t)$, related to $\sigma_J(t)$ by 
\begin{equation*}
\sigma_J(t) = t \dfrac{d}{dt} \ln \tau_J(t) ~~~ (J = {\rm III}_i), ~~~~~~~~~
\sigma_J(t) = \dfrac{d}{dt} \ln \tau_J(t) ~~~ (J = {\rm I}, {\rm II}, {\rm IV}). 
\end{equation*}
and by the slightly more involved relation \eqref{avogadro} for PV. It is the $\tau$-function that usually appears in physical problems because of its relation to spectral (Fredholm) determinants and conformal field theory (CFT); in fact also in our case the most natural object to consider is the $\tau$-function. 

The second important realization of Painlev\'{e} equations is via the theory of isomonodromic deformations of systems of linear ODEs (or flat connections). The linear systems of interest are 
\begin{equation}
\dfrac{d}{d z} \Psi(z) = \mathbf{A}(z) \Psi(z), \label{isomon}
\end{equation}
where $z$ is the affine coordinate on $\mathcal{C}_{0,n}$ (an $n$-punctured $\mathbb{CP}^1$), $\mathbf{A}(z) \in sl(2,\mathbb{C})$ is a $2 \times 2$ traceless complex matrix of meromorphic functions and $\Psi(z) \in GL(2,\mathbb{C})$ is an invertible complex $2 \times 2$ matrix. 

Let $\{z_{\nu}\}$ denote the set of poles of ${\bf A}(z)dz$. It may be assumed without loss in generality that the point at infininty does not belong to it, in which case we can write
 \begin{equation}
 \mathbf{A}(z)=\sum_{\nu=1}^n\sum_{k=1}^{r_{\nu}+1}\frac{A_{\nu,-k+1}}{
 (z-z_{\nu})^{k}}.
 \end{equation}
 The non-negative integer number $r_{\nu}$ is the Poincar\'{e} rank of the singular point $z=z_{\nu}$. If $r_{\nu}=0$, the latter is called a regular singularity where the solution $\Psi(z)$ in general develops a branch point. 
 
 The singular points with $r_{\nu}\ge1$ are irregular. The asymptotics of $\Psi(z)$ in their neighborhood exhibits the {\it Stokes phenomenon} which can be outlined as follows. Diagonalizing the highest polar contribution $A_{\nu,-r_{\nu}}=G_{\nu}\Theta_{\nu,-r_{\nu}}G_{\nu}^{-1}$, one may write a unique formal solution in the neigborhood of $z_{\nu}$ as
  \begin{equation*}
  \Psi^{(\nu)}_{\text{formal}}(z)=G_{\nu}\left[\mathbf 1+
  \sum_{k=1}^{\infty}g_{\nu,k}(z-z_{\nu})^k\right]\exp\left\{
  \Theta_{\nu,0}\ln(z-z_{\nu})+\sum_{k=1}^{r_{\nu}}
  \Theta_{\nu,-k}(z-z_{\nu})^{-k}\right\},
   \end{equation*}
 where all $\Theta_{\nu,-k}$ are diagonal. They are determined, together with
 the matrix coefficients $g_{\nu,k}$, by the linear system (\ref{isomon}). The non-formal canonical solutions $\Psi^{(\nu)}_{k}(z)$ are uniquely specified by the asymptotic condition $\Psi^{(\nu)}_{k}(z)\simeq\Psi^{(\nu)}_{\text{formal}}(z)$
 as $z\to z_{\nu}$ inside certain Stokes sectors $\mathcal S_{\nu,k}$
 centered at $z_{\nu}$.
 Canonical solutions associated to different sectors are related by $\Psi^{(\nu)}_{k+1}(z)=\Psi^{(\nu)}_{k}(z)S^{(\nu)}_k$. The Stokes matrices $S^{(\nu)}_k$, formal monodromy exponents $\Theta_{\nu,0}$ and connection matrices which relate solutions at different singular points constitute the set of generalized monodromy data of the linear system (\ref{isomon}).

One of the interesting properties of these systems of linear ODEs is that while giving a specific $\mathbf{A}(z)$ determines the global monodromy of the solution $\Psi(z)$, the converse is not true: 
a solution $\Psi(z)$ with prescribed monodromy/Stokes data 
may correspond to a family of $\mathbf{A}(z;\{\vec{t}\})$ parameterized by the moduli $\{\vec{t}\}$ of a flat $SL(2,\mathbb{C})$ bundle over the punctured sphere\footnote{One way to describe $\{\vec{t}\}$ is to combine the positions $\{z_{\nu}\}$ of singular points with diagonal elements of $\Theta_{\nu,-r_{\nu}},\ldots,\Theta_{\nu,-1}$.}. In the case we are interested in, namely a sphere
with four regular punctures and its degenerations by collisions, there is a one-parameter ($t$) family of different $\mathbf{A}(z,t)$ associated to it. One is therefore led to consider isomonodromic deformations of $\mathbf{A}(z,t)$, i.e. $t$-dependent deformations of $\mathbf{A}(z)$ which preserve the generalized monodromy of $\Psi(z,t)$; requiring the deformation $t$ to be isomonodromic implies that $\Psi(z,t)$ satisfies the system
\renewcommand\arraystretch{1.3}
\begin{equation}
 \left\{
\begin{array}{l}
\partial_z \Psi(z,t) = \mathbf{A}(z, t) \Psi(z,t), \\
\,\partial_t \Psi(z,t) = \mathbf{B}(z, t) \Psi(z,t),
\end{array} \right. \label{isosys}\renewcommand\arraystretch{1}
\end{equation}

\noindent where $\mathbf{B}(z,t)$ can be algorithmically expressed in terms of  $\mathbf{A}(z, t)$. This is an overdetermined system whose compatibility condition 
\begin{equation}
\Psi_{z t}(z,t) = \Psi_{t z}(z,t)
\end{equation}
implies the equation
\begin{equation}
\mathbf{A}_t(z, t) = \mathbf{B}_{z}(z, t) + [\mathbf{B}(z, t), \mathbf{A}(z, t)] \;\;\;
\label{cc}
\end{equation}
which yields a system of non-linear ODEs leading to the Painlev\'{e} equations; the matrices $\mathbf{A}$, $\mathbf{B}$ are known as the \textit{Lax Pair} for the associated Painlev\'{e} equation. Notice that \eqref{cc} can be seen as a gauge transformation of the flat connection $\mathbf{A}(z, t)$. 

Isomonodromic deformations admit an interesting limit to isospectral deformation problems (see for example \cite{Levin:1997tb}). In order to see this it is convenient to rescale the parameters of the problem in order to explicitly introduce a ``Planck constant'' $\kappa$ so that \eqref{isomon} becomes
\begin{equation}
(\kappa\, \partial_z - \mathbf{A}(z))\Psi(z) = 0. \label{rescaled}
\end{equation}
If we now rescale time according to $t = T_0 + \kappa\, T$ and send $\kappa \rightarrow 0$ (i.e. we focus on the dynamics around $T_0$) the connection $\kappa\, \partial_z - \mathbf{A}(z)$ reduces to a one-form $\mathbf{A} \in \Omega(C_{0,n}, sl(2,\mathbb{C}))$, that is a Higgs field in terms of Hitchin integrable systems; moreover in this limit the Hamiltonian becomes time-independent and the isomonodromic deformations reduce to isospectral deformations preserving the spectrum (eigenvalues) of $\mathbf{A}$. This can also be rephrased by saying that in this limit the spectral curve
\begin{equation}
\Sigma: ~~~ \text{det}(y - \mathbf{A}) = 0\,, ~~~ ~~~ ~~~ \Sigma \in T^*\mathcal{C}_{0,n} \label{Psc}
\end{equation}
remains fixed under the deformation; notice that for $\mathbf{A} \in sl(2,\mathbb{C})$ the spectral curve $\Sigma$ is simply
\begin{equation}
\Sigma: ~~~ y^2 = \dfrac{1}{2} \text{Tr} \mathbf{A}^2.
\end{equation}
The inverse limit, that is recovering the isomonodromic problem from the isospectral one, requires to consider Whitham deformations of the spectral curve: see for example \cite{Krichever:1992qe} or \cite{Nakatsu:1995bz,Itoyama:1995nv,Gorsky:1998rp} for a discussion of Whitham deformations in Seiberg-Witten theory.

Postponing to the next Sections the explicit expressions for the spectral curves associated to our Painlev\'{e} equations, we anticipate here their classification in terms of type of punctures\footnote{Here the degree of the puncture is the order of a pole appearing in $\frac{1}{2} \text{Tr} \mathbf{A}^2 $ and not the degree $r_{\nu}$ of $\mathbf{A}$ which appears in \eqref{isomon}. In particular, a generic regular singularity has degree 2.}:
    \begin{itemize}
    \item[] PIII$_3$: two irregular punctures of degree 3,
    \item[] PIII$_2$: two irregular punctures of degree 3 and 4, 
    \item[] PIII$_1$: two irregular punctures of degree 4,   
    \item[] PV$_{deg}$: two regular punctures and one irregular of degree 3, 
    \item[] PV: two regular punctures and one irregular of degree 4,
    \item[] PVI: four regular singularities
    \end{itemize}
and
    \begin{itemize}
    \item[] PI: one irregular puncture of degree 7, 
    \item[] PII$_{JM}$: one irregular puncture of degree 8, 
    \item[] PII$_{FN}$: one regular and one irregular puncture of degree 5, 
    \item[] PIV: one regular and one irregular puncture of degree 6.
    \end{itemize}
As we will see shortly, it is from the realization in terms of flat connections and Hitchin systems that we can understand how and why Painlev\'{e} equations are related to four-dimensional $\mathcal{N} = 2$ theories. 

\subsection{Hitchin systems and four-dimensional $\mathcal{N} = 2$ theories} 
\label{subsec:Hitchin}

The relation between Hitchin systems and four-dimensional $\CN=2$ theories in class $\CS$ is well-known in the literature \cite{Gaiotto:2009we,2009arXiv0907.3987G}. 
The idea is to consider a twisted compactification of the six-dimensional $\CN=(2,0)$ theory of type $A_1$\footnote{We will restrict here on type $A_1$ theories since this is the relevant case for Painlev\'{e} equations.} on a Riemann surface $\mathcal{C}_{g,n}$; this procedure generates a class of $\CN=2$ theories in four dimensions known as class $\mathcal{S}$ theories. Hitchin systems arise when we further compactify on a circle $S^1_R$: the low-energy effective three-dimensional theory obtained in this way is an $\mathcal{N} = 4$ sigma model with target space $\mathcal{M}$ which is  hyperK\"{a}hler and coincides with the moduli space of solutions to a Hitchin system. The appeareance of a Hitchin system can be better understood by reversing the order of compactification: first we compactify on $S^1_R$ to obtain a pure $SU(2)$ five-dimensional $\mathcal{N} = 2$ theory; the following twisted compactification on $\mathcal{C}_{g,n}$ leads to BPS equations known to be Hitchin equations. Explicitly, $\mathcal{M}$ is the moduli space of solutions $(A, \varphi)$ (with prescribed boundary conditions at the punctures of $\mathcal{C}_{g,n}$) to the Hitchin equations 
\begin{equation}
\begin{split}
F_{z \bar{z}} + R^2[\varphi_{z}, \bar{\varphi}_{\bar{z}}] &= 0, \\
\partial_{\bar{z}} \varphi_z + [A_{\bar{z}},\varphi_z] & = 0, \\
\partial_{z} \bar{\varphi}_{\bar{z}} + [A_{z},\bar{\varphi}_{\bar{z}}] & = 0,
\end{split} \label{hitchineq}
\end{equation}
modulo gauge transformations; here $z$ is a complex coordinate on $\mathcal{C}_{g,n}$, $F_{z \bar{z}}$ is the fieldstrength of the components $(A_{z}, A_{\bar{z}})$ of the five-dimensional $SU(2)$ gauge field on $\mathcal{C}_{g,n}$, while $\varphi_z$ is a complex adjoint scalar field built out of two of the five adjoint scalars of the five-dimensional theory and $\bar{\varphi}_{\bar{z}}$ its Hermitian conjugate ($\varphi_z$ is actually a (1,0)-form on $\mathcal{C}_{g,n}$ after the twisting).

The Hitchin moduli space $\mathcal{M}$ associated to $\mathcal{C}_{g,n}$ can also be thought as a torus fibration over  the moduli space $\mathcal{B}$ of Coulomb branch vacua of the $\mathcal{N} = 2$ four-dimensional theory, with generic fiber a torus of dimension dim($\mathcal{B}$). As such, it is expected to contain information on the Coulomb branch of the four-dimensional theory. This can be made quantitative in the following way.
The moduli space $\mathcal{M}$ has a hyperK\"{a}hler structure, which implies that it admits a $\mathbb{CP}^1[\zeta]$ worth of complex structures $J^{(\zeta)}$. Let $E$ denote an $SU(2)$-bundle on $\mathcal{C}_{g,n}$; when $\zeta \in \mathbb{C}^{\times}$ we are in the complex structure usually known as $J$ and we can construct a complex $E_{\mathbb{C}}$ connection $\nabla = \partial + \mathcal{A}$ with
\begin{equation}
\mathcal{A} = \dfrac{R}{\zeta} \varphi + A + R \zeta \bar{\varphi};
\end{equation}
it is also sometimes useful to decompose $\nabla$ as
\begin{equation}
(\zeta D_z + R \varphi_z, D_{\bar{z}} + \zeta R \bar{\varphi}_{\bar{z}}). \label{dec}
\end{equation}
The connection $\nabla$ is flat because of \eqref{hitchineq} and therefore the pair $(E_{\mathbb{C}}, \nabla)$ defines a flat bundle. 
On the other hand, when $\zeta = 0$ (corresponding to the complex structure known as $I$) we remain with the pair
\begin{equation}
(R \varphi_z, D_{\bar{z}});
\end{equation}
in this case the Hitchin equations \eqref{hitchineq} imply that $\varphi_z$ is a holomorphic section of $E_{\mathbb{C}}$ with respect to the holomorphic structure defined by $D_{\bar{z}}$, which means that at $\zeta = 0$ we obtain a Higgs bundle $(E_{\mathbb{C}}, \varphi_z)$\footnote{A similar story is valid at $\zeta = \infty$ (or complex structure $K$) where we obtain an anti-Higgs bundle.} whose base space is the moduli of the quadratic differential 
\begin{equation}
\frac{1}{2}\text{Tr}(\varphi_z^2) = \phi_2 \label{pororofriend}
\end{equation}
on $\mathcal{C}_{g,n}$. It is in this complex structure $I$ that the relation between Hitchin systems and Coulomb branch of four-dimensional $\mathcal{N} = 2$ theories is more manifest: in fact to every Higgs bundle one can associate a classical integrable system, whose spectral curve coincide with the Seiberg-Witten curve 
    \bea
\Sigma_{\text{SW}}: ~~~    \text{Det}(y - \varphi_z) = 0 \;\;\; \Longrightarrow \;\;\;
    y^2
     =     \phi_2 \label{SWcurve}
    \eea
of the four-dimensional theory corresponding to the chosen Riemann surface $\mathcal{C}_{g,n}$. 
Although the correspondence applies to arbitrary punctured Riemann surfaces, in order to compare with Painlev\'{e} equations we take $\mathcal{C}_{g,n}$ to be the Riemann sphere $\mathcal{C}_{0,n}$ with four punctures and its degenerations.      
The four-dimensional theories obtained by the compactification on $\mathcal{C}_{0,n}$ are $\CN=2$ $SU(2)$ gauge theories with $N_f$ fundamental hypermultiplets ($N_f\le4$), and the rank-one SCFTs of Argyres-Douglas type $H_0$, $H_1$, $H_2$ \cite{Argyres:1995jj,Argyres:1995xn} which are the maximal conformal points of the $SU(2)$ theory with $N_f=1,2,3$.
These are classified by the type of punctures as follows:
    \begin{itemize}
    \item[] $N_f = 0$: two irregular punctures of degree 3,
    \item[] $N_f = 1$: two irregular punctures of degree 3 and 4,
    \item[] $N_f = 2$: two irregular punctures of degree 4,  ~~(first realization)
    \item[] $N_f = 2$: two regular and one irregular punctures of degree 3 ,  ~~(second realization) 
    \item[] $N_f = 3$: two regular and one irregular puncture of degree 4,
    \item[] $N_f = 4$: four regular singularities
    \end{itemize}
for the $SU(2)$ theories with $N_f$ hypermultiplets, and
    \begin{itemize}
    \item[] $H_0$: one irregular puncture of degree 7,
    \item[] $H_1$: one irregular puncture of degree 8, ~~(first realization),
    \item[] $H_1$: one regular and one irregular puncture of degree 5, ~~~(second realization)
    \item[] $H_2$: one regular and one irregular puncture of degree 6
    \end{itemize}
for the Argyres-Douglas theories. 
Note that we have two realizations of the $SU(2)$ theory with $N_f=2$.
Indeed the quadratic differentials of them are different as we will see shortly, however the physical quantities computed from them are the same. Similarly there are two realizations of the $H_1$ SCFT. Let us also remark that the SW curves of the theories we are considering can be obtained from the $SU(2)$ $N_f = 4$ one by taking appropriate limits and scaling of the parameters; the web of relations between SW curves is exactly the same as the Painlev\'{e} coalescence diagram of Figure \ref{coalescence}.

\subsection{Correspondence Painlev\'{e} equations $-$ four-dimensional $\mathcal{N} = 2$ theories}
We now state the correspondence between Painlev\'{e} Lax pairs and the Hitchin systems associated to four-dimensional $\mathcal{N} = 2$ theories. The common singularity pattern on the Riemann sphere appearing both in Painlev\'{e} isomonodromy problems and M-theory compactifications naturally suggests the relation displayed in Table \ref{table:connection}.

\begin{table}[h] 
\begin{equation*}
\begin{array}{|c|c|c|c|c|c||c|c|c|c|} 
\hline
{\rm VI} & {\rm V} & {\rm III}_1 & {\rm V}_{deg} & {\rm III}_2 & {\rm III}_3 & {\rm IV} & {\rm II_{JM}} & {\rm II_{FN}} & {\rm I} \\ \hline \hline
N_f=4 & N_f=3 & N_f=2 \,(1st) & N_f=2 \,(2nd) & N_f=1 & N_f=0 & H_2 & H_1 \,(1st) & H_1 \,(2nd) & H_0 \\
\hline
\end{array}  
\end{equation*}
\caption{Correspondence between Painlev\'{e} equations and $\CN=2$ theories in four dimensions.} 
\label{table:connection} 
\end{table}

Another piece of evidence for the matching in Table \ref{table:connection} comes from considering the Higgs bundle limit of the two sides of the correspondence. As we saw, both the Painlev\'{e} connection $\kappa \partial_z - \mathbf{A}$ and the Hitchin connection $\nabla$ reduce to a Higgs bundle in the limits $\kappa \rightarrow 0$ and $\zeta \rightarrow 0$ respectively. In this limit we can compare the spectral curve relative to the Painlev\'{e} Higgs field $\mathbf{A}$ \eqref{Psc} with the Seiberg-Witten curve relative to the Hitchin Higgs field $\varphi_z$ \eqref{SWcurve}. The Seiberg-Witten curves of the theories of interest are well known: in an appropriate parameterization, the quadratic differentials $\phi_2$ of the $SU(2)$ gauge theory with $N_f =  0,1,2,3$ are written as
    \begin{equation}
    \begin{split}
    N_f = 0: & \;\; \dfrac{\Lambda^2}{z^3} + \dfrac{2u}{z^2} + \dfrac{\Lambda^2}{z}, \\
    N_f = 1: & \;\; \dfrac{\Lambda^2}{z^3} + \dfrac{3u}{z^2} + \dfrac{2\Lambda m}{z} + \Lambda^2,  \\
    N_f = 2: & \;\; \dfrac{\Lambda^2}{z^4} + \dfrac{2\Lambda m_1}{z^3} + \dfrac{4u}{z^2} + \dfrac{2\Lambda m_2}{z} + \Lambda^2,  ~~~~(\text{first realization}) \\
    N_f = 2: & \;\; \dfrac{m^2_+}{z^2} + \dfrac{m^2_-}{(z-1)^2} + \dfrac{\Lambda^2 + u}{2z} + \dfrac{\Lambda^2 - u}{2(z-1)},  ~~~~(\text{second realization}) \\
    N_f = 3: & \;\; \dfrac{m^2_+}{z^2} + \dfrac{m^2_-}{(z-1)^2} + \dfrac{2\Lambda m + u}{2z} + \dfrac{2\Lambda m - u}{2(z-1)} + \Lambda^2.
    \end{split}
    \end{equation} 
while for the Argyres-Douglas theories we have
    \begin{equation}
    \begin{split}
    H_0: & \;\; z^3 - 3 c z + u \\
    H_1: & \;\; z^4 + 4 cz^2 + 2 mz + u, \;\;\;(\text{first realization}) \\
    H_1: & \;\; z + c + \dfrac{u}{z} + \dfrac{m^2}{z^2},  \;\;\;(\text{second realization}) \\
    H_2: & \;\; z^2 + 2cz + (2\tilde{m} + c^2) + \dfrac{u + 2 c m_-}{z} + \dfrac{m_-^2}{z^2}.
    \end{split}
    \end{equation}
On the other hand, the matrix $\mathbf{A}$ is explicitly known for all Painlev\'{e} equations \cite{JM2}; in the following sections we will present these $\mathbf{A}$ and show that the associated spectral curves exactly coincide with the gauge theory ones according to Table \ref{table:connection} after an appropriate identification of parameters.
For the moment we just remark that the number of mass parameters in our four-dimensional $\mathcal{N} = 2$ theories coincide with the number of free parameters of the associated Painlev\'{e} equations (see Table \ref{numpar}) and that the Painlev\'{e} $\sigma$-function (or Hamiltonian) will turn out to correspond to the Coulomb branch parameter $u$.
We are thus led to the identification
\bea
    \frac{1}{2} \Tr(\mathbf{A})^2 = \frac{1}{2} \Tr (\varphi_z^2) = \phi_2 
\eea
from the Higgs bundle limit.

We are left with the problem of interpreting the Painlev\'{e} connection $\kappa \partial_z - \mathbf{A}$ at generic $\kappa$ in terms of Hitchin system quantities. Our Painlev\'{e} connection is holomorphic in $z$ and only involves the Higgs field; something similar exists in the Hitchin systems literature and corresponds to the Hitchin connection $\nabla$ in the \textit{oper} limit \cite{Gaiotto:2014bza}. The oper limit consists of sending $R \rightarrow 0$, $\zeta \rightarrow 0$ in such a way as to keep the ratio $\zeta/R = \hbar$ constant, so that
\begin{equation}
\nabla \;\;\; \underset{oper}{\longrightarrow} \;\;\; \hbar \partial_z + \varphi_z \;\;\; \Longleftrightarrow\;\;\; 
\kappa \partial_z - \mathbf{A}.
\end{equation}
Then the parameters $\kappa$ and $\hbar$ play the same role, and when they are sent to zero both the isomonodromic problem and the oper reduce to a Higgs bundle, while the connection of the isomonodromic problem $\mathbf{A}$ still corresponds to the Higgs field $\varphi_z$. With this identification one can think of the compatibility condition \eqref{cc} as a gauge transformation for the oper, while isomonodromic deformations \eqref{isosys} translate into the so-called Whitham deformations of the Hitchin system \cite{Krichever:1992qe,Gorsky:1998rp,Levin:1997tb}.  

To summarize, from what said previously and from the analysis we will carry out in the next sections we can construct the dictionary in Table \ref{dictionary}.

\renewcommand{\arraystretch}{1.5}
\begin{table}[h] 
\centering
\begin{tabular}{|c|c|} 
\hline
\textbf{Painlev\'{e} isomonodromic problem} & \textbf{$\mathcal{N} = 2$ theory Hitchin system} \\ \hline \hline
punctured Riemann sphere $\mathcal{C}_{0,n}[z]$ & Gaiotto surface $\mathcal{C}_{0,n}[z]$ \\ \hline
connection $\kappa \partial_z - \mathbf{A}$ & oper $\hbar \partial_z + \varphi_z$  \\ \hline
isomonodromic deformations & Whitham deformations \\ \hline
compatibility condition \eqref{cc} & gauge transformation \\ \hline
overall scale $\kappa$ & oper parameter $\hbar$ \\ \hline
isospectral limit $\kappa \rightarrow 0$ (Higgs bundle) & complex structure $I$ limit $\hbar \rightarrow 0$ (Higgs bundle) \\ \hline
Painlev\'{e} time $t$ & gauge coupling $\Lambda$ or $c$ \\ \hline
Painlev\'{e} $\sigma$-function (Hamiltonian) & Coulomb branch parameter $u$ \\ \hline
Painlev\'{e} free parameters & masses $\mathcal{N} = 2$ theory  \\ \hline
\end{tabular} 
\caption{Correspondence between Painlev\'{e} isomonodromic problems and $\mathcal{N} = 2$ theory Hitchin systems.}  
\label{dictionary} 
\end{table} 
\renewcommand{\arraystretch}{1}

\subsection{Painlev\'{e} $\tau$-functions and ``dual'' instanton partition functions}

In the previous section we saw how Painlev\'{e} equations are related to four-dimensional $\mathcal{N} = 2$ theories. 
Given this connection, one can wonder whether there are gauge theory quantities which have an analogue in Painlev\'{e} theory. In fact, an example is already known in the literature: starting with the work \cite{Gamayun:2012ma}, it turned out that Painlev\'{e} $\tau$-functions have a clear interpretation in gauge theory as the partition function (or better, the so-called ``dual'' instanton partition function). Scope of this section is to quickly review this correspondence; the reader is referred to \cite{Gamayun:2012ma} and subsequent works \cite{2013JPhA...46G5203G,Iorgov:2014vla,
Bershtein:2014yia,Gavrylenko:2016zlf} (see also \cite{moore1990,Gavrylenko:2016moe}) for further details. \\

The starting point is the work \cite{sato1977}: in that paper the authors considered the Fuchsian system with $n$ regular singularities,
\begin{equation}
\dfrac{d}{dz} \Psi(z) = \mathbf{A}(z) \Psi(z) \,,~~~~~~~~~ \mathbf{A}(z) = \sum_{\nu = 1}^{n} \dfrac{A_{\nu}}{z - z_{\nu}}
\label{ls}
\end{equation} 
 and constructed the solution $\Psi(z)$ to the isomonodromic deformation problem in terms  of free fermions; here the deformation parameters correspond to the positions of singularities. More precisely, what they were able to show is that the fundamental matrix  $\Psi(z)$ can be represented as
\begin{equation}
\Psi_{\alpha \beta}(z;\{z_{\nu}\}) = (z - z_0) \dfrac{\langle\mathcal{O}_{L_1}(z_1)\ldots\mathcal{O}_{L_{n}}(z_n) \bar{\psi}_{\alpha}(z_0) \psi_{\beta}(z) \rangle}{\langle \mathcal{O}_{L_1}(z_1)\ldots\mathcal{O}_{L_{n}}(z_n)\rangle} \label{solCFT}
\end{equation}
$(\alpha, \beta = 1,2)$ while the associated $\tau$-function can be expressed as the correlator
\begin{equation}
\tau(\{z_{\nu}\}) = \langle \mathcal{O}_{L_1}(z_1)\ldots\mathcal{O}_{L_{n}}(z_n)\rangle \label{tauCFT}
\end{equation}
The notation is as follows: the matrices $A_{\nu}$ are assumed to be diagonalizable with diagonal form $\Theta_{\nu}$; the monodromy matrices are then $M_{\nu} = C_{\nu}^{-1} e^{2\pi i \Theta_{\nu}} C_{\nu}$ with $C_{\nu}$ connection matrix, and we also introduced the logarithms of monodromy matrices, $L_{\nu} = C_{\nu}^{-1} \Theta_{\nu} C_{\nu}$. The fields $\mathcal{O}_{L_{\nu}}$ and $\bar{\psi}_{\alpha}$, $\psi_{\beta}$ are primary fields in a two-dimensional $c=1$ CFT with dimensions $\Delta_{\nu} = \frac{1}{2} \text{Tr} A_{\nu}^2$ and $\Delta_{\psi} = \frac{1}{2}$ respectively; more specifically,  $\psi$, $\bar\psi$ are free complex fermions. 

The basic idea behind the expressions \eqref{solCFT} and \eqref{tauCFT} is that the solution $\Psi(z)$ is uniquely determined by its normalization, monodromy and singular behaviour; if one can construct  $\psi$, $\mathcal{O}$ with appropriate braiding properties, then our ansatz \eqref{solCFT} must be a solution to \eqref{ls}. Normalization $\Psi(z\to z_0) = \mathbf{1}$ of the solution is guaranteed by the leading OPE term
\begin{equation}
\bar{\psi}_{\alpha}(z_0) \psi_{\beta}(z) \;\;\sim\;\; \dfrac{\delta_{\alpha \beta}}{z - z_0}
\end{equation}
while the request of having monodromy matrices $M_{\nu}$ implies that  we should look for fields $\mathcal{O}_{L_{\nu}}$ with leading OPE
\begin{equation}
\mathcal{O}_{L_{\nu}}(z_{\nu}) \psi_{\alpha}(z) \;\;\sim\;\; (z - z_{\nu})^{L_{\nu}} \mathcal{O}^{(0)}_{L_{\nu}, \alpha}(z_{\nu})
\end{equation}
for some local field $\mathcal{O}^{(0)}_{L_{\nu}, \alpha}$. An attempt of explicit construction of the fields $\mathcal{O}_{L_{\nu}}$ with such properties\footnote{These fields are known as holonomic fields, or twist fields, or spin fields. See \cite{Cecotti:1992vy} for a discussion on the relation between spin fields in Ising model and $\tau$-functions in terms of the ``new supersymmetric index''.} has been made in \cite{sato1977} and recently fully developed in \cite{Gavrylenko:2016moe}; a discussion of a similar construction in the irregular case can be found in \cite{moore1990}. 

Let us now focus on the $\tau$-function \eqref{tauCFT} in the Painlev\'{e} VI case, where $n=4$ and we can set $(z_1,z_2,z_3,z_4)=(0,t,1,\infty)$ by means of a M\"obius transformation. Let $\pm \theta_{\nu}$ be the local monodromy exponents (eigenvalues of $A_{\nu}$), so that $\Delta_{\nu} = \theta_{\nu}^2$. We are then left with the computation of a particular four-point correlator of primary fields in a $c=1$ CFT; thanks to the AGT correspondence \cite{Alday:2009aq} we know how to express this correlator via the unrefined ($\epsilon_1 = - \epsilon_2 = \epsilon$) instanton partition function of $\mathcal{N} = 2$ $SU(2)$ theory with $N_f = 4$, but there is a subtlety to be taken into account. The problem is the following: in the computation of the correlator in the, say, $s$-channel (an expansion around $t = 0$) the dimension of all primaries in the OPE $\mathcal{O}_{L_{0}}\mathcal{O}_{L_{t}}$ will enter; these primaries in the OPE will have monodromy $M_t M_0$. Let $e^{\pm 2\pi i \sigma}$ be the eigenvalues of $M_t M_0$; clearly $\sigma$ is only defined up to addition of an integer $n \in \mathbb{Z}$, so we expect that in the OPE $\mathcal{O}_{L_{0}}\mathcal{O}_{L_{t}}$ there will be an infinite number of monodromy fields $\mathcal{O}_{L_{0t}^{(n)}}$ (all possible fields with monodromy $M_t M_0$) with dimensions $(\sigma + n)^2$. With this in mind, and taking into account the small $t$ asymptotic expression for $\tau_{\text{VI}}(t)$ found in \cite{19821137}, it has been conjectured in \cite{Gamayun:2012ma} that
\begin{equation}
\tau_{\text{VI}}(t) \; \propto \; Z_{N_f = 4}^{D}(\sigma, t, \{\vec{\theta}\}) \label{DUAL}
\end{equation}
as an expansion around $t = 0$; similar expansions can also be obtained around $t = 1$ and $t = \infty$. Here $Z_{N_f = 4}^{D}$ is the so-called ``dual'' partition function for $\mathcal{N} = 2$ $SU(2)$ $N_f = 4$ theory \cite{Nekrasov:2003rj}\footnote{A similar expression also appeared as the partition function for the I-brane system of \cite{Dijkgraaf:2007sw,Aganagic:2011mi,Bonelli:2011na}.}
\begin{equation}\label{fourierNf4}
Z_{N_f = 4}^{D}(\sigma, t, \{\vec{\theta}\}) = \sum_{n \in \mathbb{Z}} e^{i n \rho} Z_{N_f = 4}(\sigma + n, t, \{\vec{\theta}\})\,.
\end{equation}
The time $t$ and the parameters $\{\vec{\theta}\}$ correspond to the energy scale $\Lambda/\epsilon$ and the masses $\{\vec{m}/\epsilon\}$ of the four flavours (here the remaining Omega background parameter $\epsilon$ is interpreted as an overall scale, analogously to the parameters $\kappa$ or $\hbar$ of Table \ref{dictionary}).
The other parameters $(\sigma, \rho)$ are integration constants for the Painlev\'{e} $\tau$-function, and we can 
think of $\sigma$ as $a/\epsilon$ in gauge theory; the interpretation of $\rho$ is less clear, although according to \cite{Nekrasov:2003rj} it should correspond to $a_D/\epsilon$ at least in the limit $\epsilon \rightarrow 0$. The proportionality factor in \eqref{DUAL} is a simple function of monodromy and $t$. 

The proposal (\ref{DUAL}) has been understood and further generalized in \cite{Iorgov:2014vla} within the framework of Liouville CFT. Crossing symmetry transformations of the Virasoro conformal blocks generate their operator-valued monodromy, which becomes particularly simple for the level 2 degenerate fields. In the latter case, the elements of monodromy matrices involve translations of intermediate Liouville momenta of conformal blocks by integer multiples of $b$ and $b^{-1}$, where $b$ parameterizes the central charge $c=1+6(b+b^{-1})^2$.  For $c=1$ such operators may be simultaneously diagonalized by
means of a Fourier type transformation, thereby combining conformal blocks with extra degenerate insertions into a solution of the Fuchsian system (\ref{ls}) with a classical $SL(2,\mathbb C)$-valued monodromy. This transformation is the origin of the summation in (\ref{fourierNf4}). 

A more direct connection to gauge theory has been established in \cite{Gavrylenko:2016zlf}. There it was shown that the tau functions of Fuchsian systems can be represented as Fredholm determinants  associated to the isomonodromic Riemann-Hilbert problem via a decomposition of the punctured Riemann sphere $\mathcal C_{0,n}$ 
into $n-2$ pairs of pants. Principal minors of this Fredholm
determinant are given by factorized expressions which reproduce Nekrasov formulas \cite{Nekrasov:2002qd,Flume:2002az,Bruzzo:2002xf,Nekrasov:2003rj} for $\mathcal{N} = 2$ $SU(2)$ $N_f = 4$ theory. 

For what other Painlev\'{e} functions are concerned, in \cite{2013JPhA...46G5203G} it was shown using the coalescence diagram in Figure \ref{coalescence} that relations similar to \eqref{DUAL} exist between the $t = 0$ expansion of the $\tau$-function for PV, PIII$_{1}$, PIII$_{2}$, PIII$_{3}$ and the dual instanton partition function expanded at weak coupling ($\Lambda = 0$) of the $\mathcal{N} = 2$ $SU(2)$ $N_f = 3, 2, 1, 0$ theories respectively. Expansions for the $\tau$-functions of PIV, PII and PI have not been systematically studied in the literature; these will be the subject of our discussion in the subsequent sections.

\begin{figure}[h]
 \centering
 \includegraphics[height=4cm]{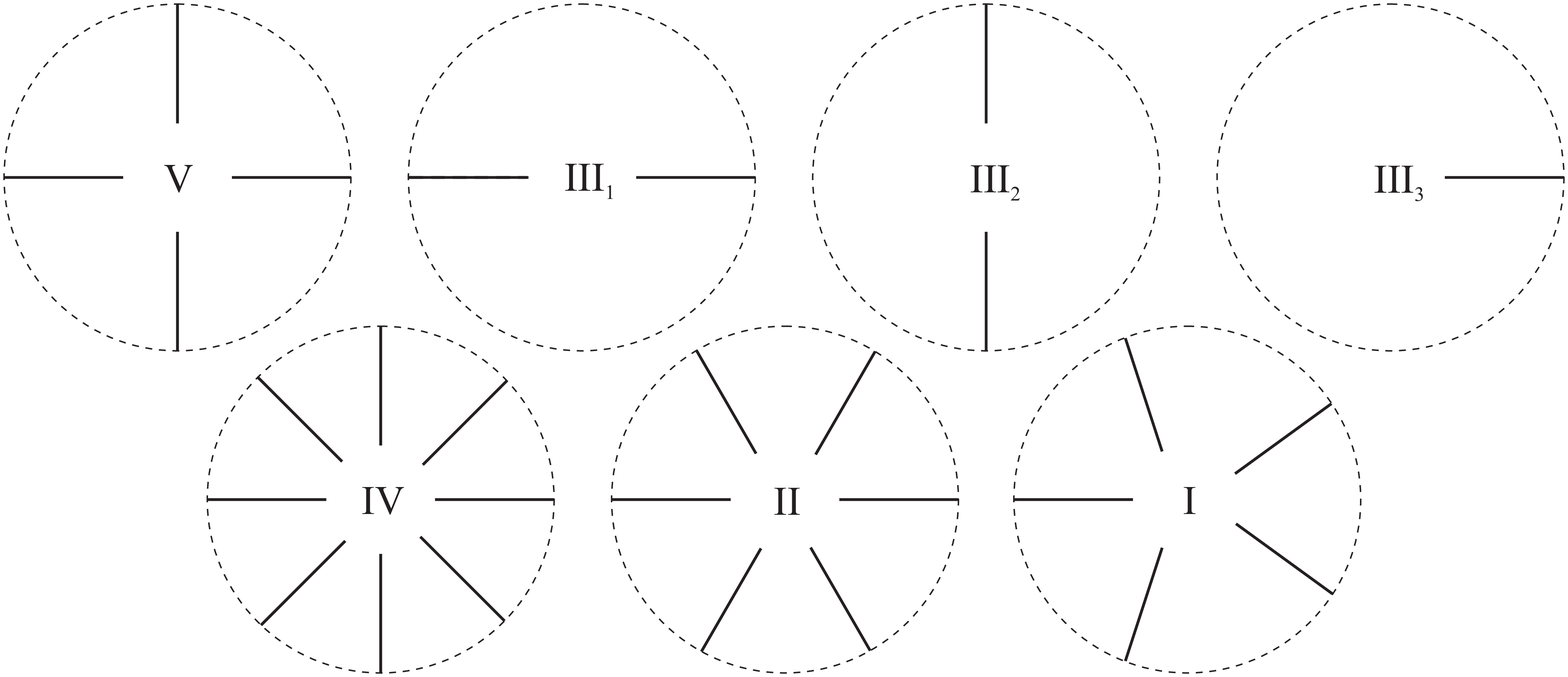}
 \caption{Canonical rays for irregular expansions at $t=\infty$.} \label{CanRays}
 \end{figure}

The first point we have to stress is that while the equations for PV, PIII$_{1}$, PIII$_{2}$, PIII$_{3}$ have two singular points at $t=0$ and $t = \infty$, PIV, PII, PI only have the singularity at $t = \infty$, cf Table~\ref{SingTable}. The expansions of Painlev\'{e} functions near $t=0$ are of regular type, whereas their irregular asymptotics as $t\to\infty$ is much richer; in particular, it presents a {\it nonlinear} Stokes phenomenon. For instance, in the case of Painlev\'{e} I, there exist 5 canonical rays $\arg t=\frac{\pi(2 k-1)}{5}$, $k=1,\ldots,5$ along which the long-distance asymptotics of $\tau_I(t)$ is trigonometric \cite{Kapaev1}, whereas inside the sectors bounded by the rays it becomes more intricate and involves elliptic functions \cite{KapaevKitaev}. In what follows, we are going to construct asymptotic expansions of the tau functions of different Painlev\'{e} equations along the canonical rays indicated in Table~\ref{tabcanrays} and  Figure~\ref{CanRays}.

\begin{table}[h] 
\centering
 \begin{tabular}{|l||c|c|c|c|c||c|c|c|}
 \hline
  & ${\rm VI}$ & ${\rm V}$ & ${\rm III}_1$ & ${\rm III}_2$ & ${\rm III}_3$ & ${\rm IV}$ & ${\rm II}$ & ${\rm I}$  \\
 \hline\hline
 $\quad$regular & $0,1,\infty$ & $0$ & $0$ & $0$ & $0$ & $-$ & $-$ & $-$ \\
 \hline
 $\quad$irregular & $-$ & $\infty$ & $\infty$& $\infty$& $\infty$& $\infty$& $\infty$& $\infty$ \\
 \hline
 \end{tabular}
 \caption{Critical points of Painlev\'{e} equations according to expansion type}
 \label{SingTable}
 \end{table}

While the correlators near $t=0$ can be easily computed via AGT since they correspond to instanton partition functions at weak coupling, correlators at $t = \infty$ would require the knowledge of the partition function at all strong coupling points (magnetic and dyonic) which we do not have\footnote{Attempts to these computations in CFT language can be found in \cite{Gaiotto:2012sf}.}. 
We will therefore proceed as follows: we make the ansatz
\begin{equation}\label{tauperiodic}
\text{``}\,\tau \; \propto \; Z^{D}\;\text{''}
\end{equation}
also along the various rays at $t=\infty$ (where this time the dual partition function is intended as computed around some strongly-coupled point\footnote{Consequently also the pair $(\sigma, \rho)$ will be matched to $a/\epsilon$, $a_D/\epsilon$ differently according to the strongly coupled point we are considering.}) and fix the coefficient of the would-be instanton expansion by requiring our $\tau(t)$ to be a solution to the $\tau$-form of the corresponding Painlev\'{e} equation. In this way we obtain putative strongly-coupled ``instanton''-like expansions at $\epsilon_1 + \epsilon_2 = 0$ for the Argyres-Douglas theories and SQCD theories; later in Section \ref{gauge} and Appendix \ref{AppB} we will check (in the limits of what is possible) that these expansions reproduce all the magnetic and dyonic strong coupling expansions of the four-dimensional theories.

\begin{table}[h] 
\centering
  \begin{tabular}{|c||c|c|c|c||c|c|c|}
  \hline
  &${\rm V}$ & ${\rm III_1}$ & ${\rm III_2}$ & ${\rm III_3}$ & ${\rm IV}$ & ${\rm II}$ & ${\rm I}$\\
  \hline\hline
  $\arg t$ & $0,\pi,\pm\frac{\pi}{2}$ & $0,\pi$ & $\pm \frac{\pi}{2}$ & $0$ & $0,\pi,\pm\frac{\pi}{4},\pm\frac{\pi}{2},\pm\frac{3\pi}{4}$
  & $0,\pi,\pm\frac{\pi}{3},\pm\frac{2\pi}{3}$ & $\pi,\pm\frac{\pi}5,
  \pm\frac{3\pi}5$\\
  \hline
  \end{tabular}
 \caption{Canonical rays at $t=\infty$.}
 \label{tabcanrays}
 \end{table}

\section{Long-distance expansions of Painlev\'{e} I, II, IV tau functions} 
\label{Painleve}

In this section we collect the long-distance expansions for the various Painlev\'{e} tau functions, along the lines of what done in \cite{2014arXiv1403.1235I} for PIII$_{3}$. 
Here we will only discuss PI, PII and PIV which only admit expansions of irregular type, while we postpone to appendix \ref{AppA} the analysis of PIII$_{1,2,3}$ and PV. These latter equations also admit regular type (short-distance) expansions around $t=0$ which can be found in \cite{2013JPhA...46G5203G}. The PVI case, in which only regular type expansions appear, has been treated at length in \cite{Gamayun:2012ma,Gavrylenko:2016zlf}.


\subsection{Painlev\'{e} I} \label{SecPI}

The Lax pair defining the monodromy preserving deformation problem \eqref{isosys} associated to PI can be found for example in \cite[eq. (C.2)]{JM2} and reads
\begin{equation}\label{PIA}
{\bf A} = A_0 +  A_1z + A_2z^2 =
\left(\begin{array}{cc}
- p & q^2 + zq + z^2 + t/2 \\ 
4z - 4q & p 
\end{array} \right) ,
\end{equation}
\begin{equation}
{\bf B} = B_0 +  B_1 z =
\left(\begin{array}{cc}
0 & \;\; q + z/2 \\ 
2 & \;\; 0 
\end{array} \right). 
\end{equation}
The compatibility condition \eqref{cc} implies that
\begin{equation}
\left\{
\begin{array}{l}
\dot{q} = p, \\
\dot{p} = 6q^2 + t, 
\end{array} \right. \label{qpI}
\end{equation}
and leads to the PI equation
\begin{equation}
\ddot{q} = 6q^2 + t.
\end{equation}
The spectral curve $y^2 = \frac{1}{2}\operatorname{Tr}{\bf A}^2$ involves the trace 
\begin{equation}
\dfrac{1}{2} \operatorname{Tr} {\bf A}^2 = 4 z^3 + 2 t z + 2 \mathcal{\sigma}_I, \label{trA2PI}
\end{equation}
where the hamiltonian function
\begin{equation}
\mathcal{\sigma}_I(t) = \dfrac{p^2}{2} - 2 q^3 - qt 
\end{equation}
satisfies the $\sigma$-PI equation
\begin{equation}\label{sigqPI}
\ddot{\sigma}_I^2 = 2 \left( \sigma_I - t \dot{\sigma}_I \right) - 4 \dot{\sigma}_I^3
\end{equation}
with respect to the dynamics \eqref{qpI}. From this one can easily extract the $\tau$-PI equation, i.e. the equation satisfied by the function $\tau_I(t)$ defined as
\begin{equation}
\sigma_I(t) = \frac{d}{dt} \ln \tau_I(t). 
\end{equation}
Different forms of PI, such as (\ref{qpI}) or (\ref{sigqPI}), possess a well-known discrete $\mathbb Z_5$-symmetry. It may be stated as follows: given a tau function $\tau_I( t)$, the transformation $\tilde\tau_I(t)=\tau_I\left(e^{\frac{2\pi i}{5}}t\right)$ yields another PI tau function.

The Stokes data describing the global asymptotic behavior of solutions
of the linear system defined by (\ref{PIA}) are parameterized by two complex quantities giving a pair of PI integrals of motion. The leading behavior of $q(t)$ as $t\to\infty$ along the canonical rays $R_k=e^{\pi i-\frac{2\pi i k}{5}}\mathbb R_{>0}$ has been described in terms
of Stokes parameters in \cite{Kapaev1}. Determining the corresponding long-distance asymptotics of $\tau_I(t)$ and using the equation  (\ref{sigqPI}) to systematically compute subleading corrections, we indeed observe a periodic pattern (\ref{tauperiodic}) similar to Painlev\'{e} VI expansion (\ref{fourierNf4}):
\\
 
\noindent \textbf{$\tau$-PI expansion} \\

\noindent On the rays arg$\,t = \pi, \pm \frac{3\pi}{5}, \pm \frac{\pi}{5}$  we have 
\begin{equation}\label{PILDE}
\begin{split}
& \tau_I(t) = s^{-\frac{1}{10}}\sum_{n \in \mathbb{Z}} e^{i n \rho} \mathcal{G}(\nu + n, s), \;\;\;\;\;\; 24t^5 + s^4 = 0,\quad s\in\mathbb R, \\
& \mathcal{G}(\nu, s) = C(\nu,s) \left[ 1 + \sum_{k=1}^{\infty} \dfrac{D_k(\nu)}{s^k} \right], \\
& C(\nu,s) = (2\pi)^{-\frac{\nu}{2}}e^{\frac{s^2}{45} + \frac{4}{5}i \nu s - \frac{i \pi \nu^2}{4}} s^{\frac{1}{12} - \frac{\nu^2}{2}} 48^{- \frac{\nu^2}{2}} G(1+\nu),
\end{split}
\end{equation} 
with $G(1 + \nu)$ Barnes $G$-function. These functions would correspond to the 1-loop part of our strong-coupling ``Nekrasov-like'' partition function $\mathcal{G}(\nu, s)$, and from them we can read how many light particles there are in the strongly coupled sector under consideration: in the case at hand there is just one Barnes $G$-function and therefore a single particle is light. On the other hand, the coefficients $D_k(\nu)$ would correspond to the ``$k$-th instanton'' contribution to the partition function; they can be computed recursively, and the first few of them are explicitly given by
\begin{equation}
\begin{split}
& D_1(\nu) = - \dfrac{i \nu (94 \nu^2 + 17)}{96}, \\
& D_2(\nu) = - \dfrac{44180 \nu^6 + 170320 \nu^4 + 74985 \nu^2 + 1344}{92160}.
\end{split}
\end{equation}
Different rays are characterized by different expressions of the integration constants $\nu,\rho$ in terms of the Stokes data, which  is a signature of  the nonlinear Stokes phenomenon.

It is plausible that the Fourier representation of $\tau_I(t)$ given by the 1st line of (\ref{PILDE}) remains valid in the whole complex plane of PI  variable~$t$. The 2nd line should then be interpreted as an asymptotic expansion of the function $\mathcal G(\nu,s)$ (PI irregular conformal block) as $s\to\infty$. The latter expansion may be expected to hold inside certain sectors; specific rays $R_k$ are singled out by the condition for the Fourier sum to be a well-defined asymptotic series for the tau function.

In order to later make contact with four-dimensional theories we do the following: we redefine $s \rightarrow s/\epsilon$, $\nu \rightarrow \nu/\epsilon$ and group the terms with the same power of $\epsilon$ in $\mathcal{G}(\nu/\epsilon, s/\epsilon)$. By using 
\begin{equation}
\ln G(1 + \nu) = \left( \dfrac{\nu^2}{2} - \dfrac{1}{12} \right) \ln \nu - \dfrac{3}{4} \nu^2 + \dfrac{\nu}{2} \ln 2\pi + \zeta'(-1) + \sum_{k \geqslant 1} \dfrac{B_{2k+2}}{4k(k+1)\nu^{2k}},
\end{equation}
it is easy to show that the logarithm of the function $\mathcal{G}(\nu/\epsilon, s/\epsilon)$ nicely reorganizes into a genus expansion, which as we will see is very natural in gauge theory if we interpret $\mathcal{G}(\nu/\epsilon, s/\epsilon)$ as the partition function for a four-dimensional $\mathcal{N}=2$ theory in the special Omega background $\epsilon_1 = - \epsilon_2 = \epsilon$. More in detail we obtain
\begin{equation}
\ln \left[ \mathcal{G}\left(\frac{\nu}{\epsilon}, \frac{s}{\epsilon}\right) \right] \;=\;
\sum_{g \geqslant 0} \epsilon^{2g-2} \mathcal{F}_g(\nu, s), \label{exI1}
\end{equation}
with 
\begin{equation}
\begin{split}
\mathcal{F}_0(\nu, s) & = \dfrac{s^2}{45} + \dfrac{4}{5} i \nu s + \dfrac{\nu^2}{2} \ln \dfrac{\nu}{48i s} - \dfrac{3\nu^2}{4} - \dfrac{47 i \nu^3}{48 s} - \dfrac{7717 \nu^4}{4608 s^2} + O(s^{-3}), \\
\mathcal{F}_1(\nu, s) & = \zeta'(-1) - \dfrac{1}{12}\ln \dfrac{\nu}{s} - \dfrac{17 i \nu}{96 s} - \dfrac{3677 \nu^2}{4608 s^2} + O(s^{-3}), \\
\mathcal{F}_2(\nu, s) & = - \dfrac{1}{240 \nu^2} - \dfrac{7}{480 s^2} + O(s^{-3}) .
\label{PIlogexp}
\end{split}
\end{equation}
and similarly for higher genus $\mathcal{F}_g(\nu, s)$ functions.

\subsection{Painlev\'{e} II} \label{SecPII}
There exist several realizations of PII in terms of isomonodromy problems. The Jimbo-Miwa Lax pair is given by \cite[eq. (C.10)]{JM2} 
\begin{equation}
{\bf A} = A_0 + A_1z + A_2z^2 =
\left(\begin{array}{cc}
z^2 + p + t/2 & u (z - q) \\ 
- \frac{2}{u} (p z + \theta + p q)\; & \;-(z^2 + p + t/2) 
\end{array} \right), 
\end{equation}
\begin{equation}
{\bf B} = B_0 +  B_1z =
\left(\begin{array}{cc}
z/2 & u/2 \\ 
- p/u \; & \; -z/2 
\end{array} \right),
\end{equation}
and corresponds to a single irregular puncture of degree $8$ at infinity. The compatibility condition \eqref{cc} requires
\begin{equation}
\left\{
\begin{array}{l}
\dot{u} = -q u, \\
\dot{q} = p + q^2 + t/2, \\
\dot{p} = - 2 p q - \theta ,
\end{array} \right. \label{qpII}
\end{equation}
and therefore implies the PII equation
\begin{equation}
\ddot{q} = 2q^3 + qt + \left(\frac{1}{2} - \theta\right). \label{PII}
\end{equation}
From the trace 
\begin{equation}
\dfrac{1}{2} \operatorname{Tr} {\bf A}^2 = z^4 + t z^2 - 2 \theta z + 2 \sigma_{II} + \dfrac{t^2}{4} \label{trA2PII}
\end{equation}
one may recover the hamiltonian function $\sigma_{II}(t) = \frac{d}{dt} \ln \tau_{II}(t)$  
\begin{equation}
\mathcal{\sigma}_{II}(t) = \dfrac{p^2}{2} + p q^2 + \dfrac{p t}{2}  + q \theta,
\end{equation}
which satisfies the $\sigma$-form of PII:
\begin{equation}
\ddot{\sigma}_{II}^2 = 2 \dot{\sigma}_{II} \left( \sigma_{II} - t \dot{\sigma}_{II} \right) - 4 \dot{\sigma}_{II}^3 + \dfrac{\theta^2}{4}. \label{sPII}
\end{equation}
 Similarly to Painlev\'{e} I, the PII equation has a discrete $\mathbb Z_3$-symmetry acting on the tau functions as $\tilde\tau_{II}(t)=\tau_{II}\left(e^{\frac{2\pi i}{3}}t\right)$.

We could also have started from the Flaschka-Newell Lax pair, describing isomonodromic deformations for systems with one regular puncture and an irregular puncture of degree~$5$. It can be written as \cite{Flaschka}
\begin{equation}
{\bf A} = A_0z^{-1} + A_1 +  A_2z = \left(\begin{array}{cc}
\frac{p q}{z} & \frac{p^2q^2 - \frac{\theta_0^2}{4}}{q z} + q + t + z \\ 
1-\frac{q}{z} & -\frac{p q}{z}  
\end{array} \right) ,
\end{equation}
\begin{equation}
{\bf B} = B_0 + B_1z =
\left(\begin{array}{cc}
0 & t + z + 2q \\ 
1 \; & \; 0 
\end{array} \right). 
\end{equation}
The zero-curvature condition \eqref{cc} becomes
\begin{equation}
\left\{
\begin{array}{l}
\dot{q} = -2 p q \\
\dot{p} = p^2 - 2 q - t - \dfrac{\theta^2}{4q^2} 
\end{array} \right.
\end{equation}
and is equivalent to the equation
\begin{equation}
\ddot{q} = \dfrac{\dot{q}^2}{2q} + 4q^2 + 2qt - \dfrac{\theta^2}{2q},
\end{equation}
which is related to PII by a change of variables. The trace
\begin{equation}
\dfrac{1}{2} \operatorname{Tr} {\bf A}^2= z + t - \frac{\sigma_{II}'}{z} + \frac{\theta_0^2}{4 z^2} \label{trA2PIIFN}
\end{equation}
contains the function  
\begin{equation}
\sigma_{II}'(t) = -\frac{p^2}{q} + q^2 + qt + \frac{\theta_0^2}{4q}
\end{equation}
satisfying
\begin{equation}
\dfrac{1}{4}\left(\ddot{\sigma'}_{II}\right)^2 = - \dot{\sigma'}_{II} (\sigma_{II}' - t \dot{\sigma'}_{II}) + \dot{\sigma'}_{II}^3 + \dfrac{\theta^2}{4}.
\end{equation}
The latter equation is related to \eqref{sPII} by rescalings $\sigma_{II}' = \sigma_{II}/2$ and $t \rightarrow -t/2$. \\

 The set of PII canonical rays listed in Table~\ref{tabcanrays} splits into two orbits of the $\mathbb Z_3$-action, cf \cite[Chapters 9-10]{Fokas:2006:PTR}. Returning to \eqref{sPII}, we accordingly find two types of long-distance expansions for the tau function $\tau_{II}(t)$ (cf \cite[Subsection 4.6]{ILP} where the special case $\theta=\frac12$ has been studied): \\

\noindent \textbf{$\tau$-PII expansion 1} \\

\noindent The asymptotics of the first type is valid on the rays arg$\,t = \pi, \pm \frac{\pi}{3}$ (i.e.  $s \to \pm i\infty$ in the formulas below). Along these rays, we have the expansion
\begin{equation}
\begin{split}
& \tau_{II}(t) = s^{-\frac{1}{6} + \frac{\theta^2}{3}} \sum_{n \in \mathbb{Z}} e^{i n \rho} \mathcal{G}(\nu + n, s), \;\;\;\;\;\; 4t^3 = 9s^2, \\
& \mathcal{G}(\nu, s) = C(\nu,s) \left[ 1 + \sum_{k=1}^{\infty} \dfrac{D_k(\nu)}{s^k} \right], \\
& C(\nu,s) = (2\pi)^{-\frac{\nu}{2}} e^{-\frac{3s^2}{32} + \nu s} s^{\frac{1}{12} - \frac{\nu^2}{2}} 12^{- \frac{\nu^2}{2}} G(1+\nu),
\end{split}
\end{equation} 
which again has a Fourier transform structure with respect to the Stokes data encoded into the complex parameters $\nu$, $\rho$. The first few expansion coefficients are given by
\begin{equation}
\begin{split}
& D_1(\nu) = \dfrac{\nu (34 \nu^2 - 96 \theta^2 + 31)}{72}, \\
& D_2(\nu) = \dfrac{289}{2592}\nu^6 - \dfrac{408 \theta^2 - 413}{648} \nu^4 + \left( \dfrac{8 \theta^4}{9} - \dfrac{187 \theta^2}{54} + \dfrac{11509}{10368} \right)\nu^2 + \dfrac{16 \theta^4 - 40 \theta^2 + 9}{216}.
\end{split}
\end{equation}
The number of Barnes functions implies that this sector contains only one light particle.
As we did for $\tau_I(t)$ we now redefine $s \rightarrow s/\epsilon$, $\nu \rightarrow \nu/\epsilon$ and $\theta \rightarrow \theta/\epsilon$ and collect the terms of the same weight in $\epsilon$ in $\mathcal{G}(\nu/\epsilon, \theta/\epsilon, s/\epsilon)$; this yields
\begin{equation}
\ln \left[ \mathcal{G}\left(\frac{\nu}{\epsilon}, \frac{\theta}{\epsilon}, \frac{s}{\epsilon} \right) \right] \;=\;
\sum_{g \geqslant 0} \epsilon^{2g-2} \mathcal{F}_g(\nu, \theta, s),
\end{equation}
with 
\begin{equation}
\begin{split}
\mathcal{F}_0(\nu, \theta, s) & = -\dfrac{3 s^2}{32} + \nu s + \dfrac{\nu^2}{2} \ln \dfrac{\nu}{12 s} - \dfrac{3\nu^2}{4} + \dfrac{17 \nu^3}{36 s} - \dfrac{4 \theta^2 \nu}{3 s} + \dfrac{125 \nu^4}{288 s^2}
- \dfrac{26 \theta^2 \nu^2}{9s^2} + \dfrac{2\theta^4}{27 s^2} + O(s^{-3}), \\
\mathcal{F}_1(\nu, \theta, s) & = \zeta'(-1) - \dfrac{1}{12}\ln \dfrac{\nu}{s} + \dfrac{31 \nu}{72 s} + \dfrac{293 \nu^2}{288 s^2} - \dfrac{5 \theta^2}{27 s^2} + O(s^{-3}), \\
\mathcal{F}_2(\nu, \theta, s) & = - \dfrac{1}{240 \nu^2} + \dfrac{1}{24 s^2} + O(s^{-3}).
\label{PIIlogexp1}
\end{split}
\end{equation}
The higher genus functions $\mathcal{F}_g(\nu, \theta, s)$ are obtained in a similar fashion. \\

\noindent \textbf{$\tau$-PII expansion 2} \\

\noindent On the complementary rays arg$\,t = 0, \pm \frac{2\pi}{3}$ (i.e. $s \to \pm\infty$ below) we have another expansion which still has Fourier type:
\begin{equation}
\begin{split}
& \tau_{II}(t) = s^{-\frac{1}{6} + \frac{\theta^2}{3}} \sum_{n \in \mathbb{Z}} e^{i n \rho} \mathcal{G}(\nu + n, s), \;\;\;\;\;\; 8t^3 = 9s^2, \\
& \mathcal{G}(\nu, s) = C(\nu,s) \left[ 1 + \sum_{k=1}^{\infty} \dfrac{D_k(\nu)}{s^k} \right], \\
& C(\nu,s) = (2\pi)^{-\nu} e^{i \nu s + \frac{i \pi \nu^2}{2}} s^{-\nu^2 + \frac{1}{6} - \frac{\theta^2}{4}} 6^{- \nu^2}G\left(1+\nu + \frac{\theta}{2}\right) G\left(1+\nu - \frac{\theta}{2}\right).
\end{split}
\end{equation} 
The first few coefficients of the sought-for PII irregular conformal block are
\begin{equation}
\begin{split}
& D_1(\nu) = - \dfrac{i \nu (68 \nu^2 - 9 \theta^2 + 2)}{36}, \\
& D_2(\nu) = -\dfrac{289}{162}\nu^6 + \dfrac{153 \theta^2 - 1159}{324} \nu^4 - \left( \dfrac{\theta^4}{32} - \dfrac{11 \theta^2}{18} + \dfrac{271}{648} \right)\nu^2 - \dfrac{\theta^2(11 \theta^2 - 68)}{1728}.
\end{split}
\end{equation}
From the number of Barnes $G$-functions we see that there is an $SU(2)$ doublet of light particles in this sector. The same procedure as above generates the genus expansion
\begin{equation}
\ln \left[ \mathcal{G}\left(\frac{\nu}{\epsilon}, \frac{\theta}{\epsilon}, \frac{s}{\epsilon} \right) \right] \;=\;
\sum_{g \geqslant 0} \epsilon^{2g-2} \mathcal{F}_g(\nu, \theta, s),
\end{equation}
where 
\begin{equation}
\begin{split}
\mathcal{F}_0(\nu, \theta, s) & = i \nu s + i \dfrac{\pi \nu^2}{2} 
+ \dfrac{(\nu + \theta/2)^2}{2} \ln \dfrac{\nu + \theta/2}{s} + \dfrac{(\nu - \theta/2)^2}{2} \ln \dfrac{\nu - \theta/2}{s} - \nu^2 \ln 6 \\
& - \dfrac{3\nu^2}{2} - \dfrac{3 \theta^2}{8} - \dfrac{17 i \nu^3}{9 s} + \dfrac{i \theta^2 \nu}{4 s} - \dfrac{125 \nu^4}{36 s^2} + \dfrac{43 \theta^2 \nu^2}{72s^2} - \dfrac{11\theta^4}{1728 s^2} + O(s^{-3}), \\
\mathcal{F}_1(\nu, \theta, s) & = 2 \zeta'(-1) - \dfrac{1}{12}\ln \dfrac{\nu + \theta/2}{s} - \dfrac{1}{12}\ln \dfrac{\nu - \theta/2}{s} - \dfrac{i \nu}{18 s} - \dfrac{5 \nu^2}{12 s^2} + \dfrac{17 \theta^2}{432 s^2} + O(s^{-3}), \\
\mathcal{F}_2(\nu, \theta,s) & = - \dfrac{1}{240 (\nu + \theta/2)^2} - \dfrac{1}{240 (\nu - \theta/2)^2} + O(s^{-3}), 
\label{PIIlogexp2}
\end{split}
\end{equation}
and similarly for higher genus $\mathcal{F}_g(\nu, \theta, s)$ functions. 


\subsection{Painlev\'{e} IV} \label{SecPIV}

The Lax pair associated to the PIV equation can be found in \cite{JM2} and reads 
\begin{equation}
{\bf A} =  \frac{A_0}{z} + A_1 + A_2z =
\left(\begin{array}{cc}
z + t + \frac{1}{z}(\theta_0 - p q) & u\left(1 - \frac{q}{2z}\right) \\ 
\frac{2}{u} (p q - \theta_0 - \theta_{\infty}) + \frac{2p}{uz}(pq-2\theta_0)\; & \; -z - t - \frac{1}{z}(\theta_0 - p q)
\end{array} \right), 
\end{equation}
\begin{equation}
{\bf B} = B_0 + B_1 z =
\left(\begin{array}{cc}
z & u \\ 
\frac{2}{u} (p q - \theta_0 - \theta_{\infty}) \; & \; -z
\end{array} \right).
\end{equation}
The compatibility condition \eqref{cc} gives
\begin{equation}
\left\{
\begin{array}{l}
\dot{u} = -(2t+q)u, \\
\dot{q} = q^2 +2q(t-2p) + 4 \theta_0, \\
\dot{p} = 2p^2 -2p(q+t) + \theta_0 + \theta_{\infty},
\end{array} \right.
\end{equation}
and implies the PIV equation
\begin{equation}
\ddot{q} = \dfrac{\dot{q}^2}{2q} + \dfrac{3q^3}{2} + 4 t q^2 + 2q(t^2 - 2\theta_{\infty} + 1) - \dfrac{8\theta_0^2}{q}. \label{PIV}
\end{equation}
The trace 
\begin{equation}
\dfrac{1}{2} \operatorname{Tr}{\bf A}^2 = z^2 + 2 z t + t^2 - 2 \theta_{\infty} + \dfrac{\sigma_{IV} + 2 t \theta_0}{z} + \dfrac{\theta_0^2}{z^2} \label{trA2PIV}
\end{equation}
involves the function $\sigma_{IV}(t) = \frac{d}{dt} \ln \tau_{IV}(t)$
\begin{equation}
\sigma_{IV}(t) = 2p^2 q - p\left( q^2 + 2 q t + 4\theta_0 \right) + q(\theta_0 + \theta_{\infty}), 
\end{equation}
satisfying the $\sigma$-PIV equation 
\begin{equation}
\ddot{\sigma}_{IV}^2 = 4 \left( t \dot{\sigma}_{IV} - \sigma_{IV} \right)^2 
- 4 \dot{\sigma}_{IV} \left(\dot{\sigma}_{IV} + 4 \theta_0 \right) \left(\dot{\sigma}_{IV} + 2 \theta_0 + 2 \theta_{\infty}\right). \label{ooo}
\end{equation}
From this one can easily deduce the $\tau$-PIV equation. We prefer to redefine $\sigma_{IV} \rightarrow \sqrt{2} \sigma_{IV}$, $t \rightarrow t/\sqrt{2}$, $\theta_0 = - \theta_s$ and $\theta_0 + \theta_{\infty} = - 2 \theta_t$ so that \eqref{ooo} becomes
\begin{equation}
\ddot{\sigma}_{IV}^2 = \left( t \dot{\sigma}_{IV} - \sigma_{IV} \right)^2 
- 4 \dot{\sigma}_{IV} \left(\dot{\sigma}_{IV} - 2 \theta_s \right) \left(\dot{\sigma}_{IV} - 2 \theta_t \right).
\end{equation}
The latter equation has a $\mathbb Z_4$-symmetry, according to which
we find two types of long-distance expansions for $\tau_{IV}(t)$
on the canonical rays: \\
 
\noindent \textbf{$\tau$-PIV expansion 1} \\

\noindent Periodic expansions of the first type are valid on four rays arg$\,t = 0, \pi, \pm \frac{\pi}{2}$ (i.e. $s \to \pm\infty$):
\begin{equation}
\begin{split}
& \tau_{IV}(t) = s^{-\frac{1}{4} + 2(\theta_s^2 - \theta_s \theta_t + \theta_t^2)}\sum_{n \in \mathbb{Z}} e^{i n \rho} \mathcal{G}(\nu + n, s), \;\;\;\;\;\; s = \dfrac{t^2}{2\sqrt{3}}, \\
& \mathcal{G}(\nu, s) = C(\nu,s) \left[ 1 + \sum_{k=1}^{\infty} \dfrac{D_k(\nu)}{s^k} \right], \\
& C(\nu,s) = (2\pi)^{-\frac{\nu}{2}} e^{\frac{s^2}{9} + i \nu s - \frac{i \pi \nu^2}{4} + \frac{2}{\sqrt{3}}(\theta_s + \theta_t)s} s^{\frac{1}{12} - \frac{\nu^2}{2}} 6^{- \frac{\nu^2}{2}} G(1+\nu),
\end{split}
\end{equation} 
where the first few coefficients are given by
\begin{equation}
\begin{split}
D_1(\nu) =& - \dfrac{i \nu^3}{3} + 2 i \nu \left( 3\left(\theta_s^2 - \theta_s \theta_t + \theta_t^2 \right) - \dfrac{1}{3} \right) - \dfrac{2(\theta_s + \theta_t)(2 \theta_s - \theta_t)(\theta_s - 2 \theta_t)}{\sqrt{3}} , \\
D_2(\nu) =& -\dfrac{\nu^6}{18} + \left( 2\left( \theta_s^2 - \theta_s \theta_t + \theta_t^2 \right) - \dfrac{67}{144} \right) \nu^4 + \dfrac{2i(\theta_s + \theta_t)(2 \theta_s - \theta_t)(\theta_s - 2 \theta_t)}{3\sqrt{3}} \nu^3 \\
& - \left( 18 \left( \theta_s^2 - \theta_s \theta_t + \theta_t^2 \right)^2 - \dfrac{29}{2} \left( \theta_s^2 - \theta_s \theta_t + \theta_t^2 \right) + \dfrac{71}{48} \right)\nu^2 \\
& - \dfrac{i(\theta_s + \theta_t)(2 \theta_s - \theta_t)(\theta_s - 2 \theta_t)(36 \left( \theta_s^2 - \theta_s \theta_t + \theta_t^2 \right) - 31)}{3 \sqrt{3}} \nu \\
& + \left( \dfrac{2}{3}(\theta_s + \theta_t)^2(2 \theta_s - \theta_t)^2(\theta_s - 2 \theta_t)^2 - 3 \left( \theta_s^2 - \theta_s \theta_t + \theta_t^2 \right)^2 + \dfrac{5}{4}\left( \theta_s^2 - \theta_s \theta_t + \theta_t^2 \right) - \dfrac{1}{12} \right).
\end{split}
\end{equation}
The number of Barnes functions implies that there is only one light particle in this sector. 
After rescaling all of the parameters by $\epsilon$, the logarithm of PIV conformal block can be expanded as
\begin{equation}
\ln \left[ \mathcal{G}\left(\frac{\nu}{\epsilon}, \frac{\theta_s}{\epsilon}, \frac{\theta_t}{\epsilon}, 
\frac{s}{\epsilon} \right) \right] \;=\;
\sum_{g \geqslant 0} \epsilon^{2g-2} \mathcal{F}_g(\nu, \theta_s, \theta_t, s),
\end{equation}
where 
\begin{equation}
\begin{split}
\mathcal{F}_0(\nu, \theta_s, \theta_t, s) & = \dfrac{s^2}{9} + i\nu s + \dfrac{2(\theta_s + \theta_t)s}{\sqrt{3}} + \dfrac{\nu^2}{2} \ln \dfrac{\nu}{6 i s} - \dfrac{3\nu^2}{4} \\
& - \dfrac{i \nu^3}{3 s} + \dfrac{6 i (\theta_s^2 - \theta_s \theta_t + \theta_t^2) \nu}{s} - \dfrac{2(\theta_s + \theta_t)(2\theta_s - \theta_t)(\theta_s - 2\theta_t)}{\sqrt{3}s} \\  
& - \dfrac{35 \nu^4}{144 s^2} + \dfrac{21 (\theta_s^2 - \theta_s \theta_t + \theta_t^2) \nu^2}{2s^2} + \dfrac{3\sqrt{3}i (\theta_s + \theta_t)(2\theta_s - \theta_t)(\theta_s - 2\theta_t) \nu}{s^2} \\
& - \dfrac{3(\theta_s^2 - \theta_s \theta_t + \theta_t^2)^2}{s^2} + O(s^{-3}), \\
\mathcal{F}_1(\nu, \theta_s, \theta_t, s) & = \zeta'(-1) - \dfrac{1}{12}\ln \dfrac{\nu}{s} - \dfrac{2 i \nu}{3 s} - \dfrac{181 \nu^2}{144 s^2} + \dfrac{5 (\theta_s^2 - \theta_s \theta_t + \theta_t^2)}{4 s^2} + O(s^{-3}), \\
\mathcal{F}_2(\nu, \theta_s, \theta_t, s) & = - \dfrac{1}{240 \nu^2} - \dfrac{1}{12 s^2} + O(s^{-3}).
\label{PIVlogexp1}
\end{split}
\end{equation}

\noindent \textbf{$\tau$-PIV expansion 2} \\

\noindent On the complementary rays arg$\,t = \pm \frac{\pi}{4}, \pm \frac{3\pi}{4}$ (i.e. $s \in i \mathbb{R}$) we have
\begin{equation}
\begin{split}
& \tau_{IV}(t) = s^{-\frac{1}{4} + 2(\theta_s^2 - \theta_s \theta_t + \theta_t^2)}\sum_{n \in \mathbb{Z}} e^{i n \rho} \mathcal{G}(\nu + n, s), \;\;\;\;\;\; s = \dfrac{t^2}{2}, \\
& \mathcal{G}(\nu, s) = C(\nu,s) \left[ 1 + \sum_{k=1}^{\infty} \dfrac{D_k(\nu)}{s^k} \right], \\
& C(\nu,s) = (2\pi)^{-\frac{3\nu}{2}} e^{\nu s + \frac{2}{3}(\theta_s + \theta_t)s} s^{-\frac{3\nu^2}{2} + \frac{1}{4} - \frac{4}{3}\left( \theta_s^2 - \theta_s \theta_t + \theta_t^2 \right)} 2^{- \frac{3\nu^2}{2}} \\ 
& \hspace{1.3 cm} G\left(1+\nu + \frac{2\theta_s - 4 \theta_t}{3}\right) G\left(1+\nu + \frac{2\theta_t - 4 \theta_s}{3}\right)G\left(1+\nu + \frac{2\theta_t + 2 \theta_s}{3}\right),
\end{split}
\end{equation} 
where the first few coefficients are 
\begin{equation}
\begin{split}
D_1(\nu) = & 3 \nu^3 - 2 \left( \theta_s^2 - \theta_s \theta_t + \theta_t^2 \right)\nu - \dfrac{2}{9}(\theta_s + \theta_t)(2 \theta_s - \theta_t)(\theta_s - 2 \theta_t), \\
D_2(\nu) = & \dfrac{9}{2}\nu^6 - \left( 6 \left( \theta_s^2 - \theta_s \theta_t + \theta_t^2 \right) - \dfrac{105}{16} \right) \nu^4 - \dfrac{2}{3} (\theta_s + \theta_t)(2 \theta_s - \theta_t)(\theta_s - 2 \theta_t) \nu^3 \\
& + \left( 2 \left( \theta_s^2 - \theta_s \theta_t + \theta_t^2 \right)^2 - \dfrac{11}{2}\left( \theta_s^2 - \theta_s \theta_t + \theta_t^2 \right) + \dfrac{3}{16} \right) \nu^2 \\
& + \dfrac{1}{9} (\theta_s + \theta_t)(2 \theta_s - \theta_t)(\theta_s - 2 \theta_t) \left( 4 \left( \theta_s^2 - \theta_s \theta_t + \theta_t^2 \right) - 7 \right) \nu \\
& + \left( \dfrac{2}{81}(\theta_s + \theta_t)^2(2 \theta_s - \theta_t)^2(\theta_s - 2 \theta_t)^2 + \dfrac{\left( \theta_s^2 - \theta_s \theta_t + \theta_t^2 \right)\left( 4 \left( \theta_s^2 - \theta_s \theta_t + \theta_t^2 \right) - 3 \right)}{36} \right).
\end{split}
\end{equation}
From the number of Barnes $G$-functions we see that there is an $SU(3)$ triplet of light particles in this sector.
Let us note that this second expansion is related to a representation of $\tau_{IV}(t)$  proposed in \cite[Conjecture 4.2]{Nagoya:2015}, \cite{Nagoya:2016}. There the function $\mathcal G(\nu,s)$ was  interpreted as $c=1$ Virasoro conformal block of a new type; it involves an irregular vertex operator intertwining rank 2 Whittaker modules. 

After rescaling all parameters by $\epsilon$ we thereby obtain
\begin{equation}
\ln \left[ \mathcal{G}\left(\frac{\nu}{\epsilon}, \frac{\theta_s}{\epsilon}, \frac{\theta_t}{\epsilon}, 
\frac{s}{\epsilon} \right) \right] \;=\;
\sum_{g \geqslant 0} \epsilon^{2g-2} \mathcal{F}_g(\nu, \theta_s, \theta_t, s),
\end{equation}
where 
\begin{equation}
\begin{split}
\mathcal{F}_0(\nu, \theta_s, \theta_t, s) & = \nu s + \frac23(\theta_s + \theta_t)s - \dfrac{3\nu^2}{2} \ln 2 - \dfrac{9\nu^2}{4} - 2(\theta_s^2 - \theta_s \theta_t + \theta_t^2) \\
& + \dfrac{(\nu + 2\frac{\theta_s - 2\theta_t}{3})^2}{2} \ln \dfrac{\nu + 2\frac{\theta_s - 2\theta_t}{3}}{s} + \dfrac{(\nu + 2\frac{\theta_t - 2\theta_s}{3})^2}{2} \ln \dfrac{\nu + 2\frac{\theta_t - 2\theta_s}{3}}{s} \\
& + \dfrac{(\nu + 2\frac{\theta_s + \theta_t}{3})^2}{2} \ln \dfrac{\nu + 2\frac{\theta_s + \theta_t}{3}}{s} \\
& + \dfrac{3\nu^3}{s} - \dfrac{2(\theta_s^2 - \theta_s \theta_t + \theta_t^2) \nu}{s} - \dfrac{2(\theta_s + \theta_t)(2\theta_s - \theta_t)(\theta_s - 2\theta_t)}{9 s} \\  
& + \dfrac{105 \nu^4}{16 s^2} - \dfrac{11 (\theta_s^2 - \theta_s \theta_t + \theta_t^2) \nu^2}{2s^2} - 
\dfrac{7 (\theta_s + \theta_t)(2\theta_s - \theta_t)(\theta_s - 2\theta_t) \nu}{9 s^2} \\ 
& + \dfrac{(\theta_s^2 - \theta_s \theta_t + \theta_t^2)^2}{9 s^2} + O(s^{-3}), \\
\mathcal{F}_1(\nu, \theta_s, \theta_t, s) & = 3\zeta'(-1) - \dfrac{1}{12}\ln \dfrac{\nu + 2\frac{\theta_s - 2\theta_t}{3}}{s} - \dfrac{1}{12}\ln \dfrac{\nu + 2\frac{\theta_t - 2\theta_s}{3}}{s} 
- \dfrac{1}{12}\ln \dfrac{\nu + 2\frac{\theta_s + \theta_t}{3}}{s} \\
& + \dfrac{3 \nu^2}{16 s^2} - \dfrac{\theta_s^2 - \theta_s \theta_t + \theta_t^2}{12 s^2} + O(s^{-3}), \\
\mathcal{F}_2(\nu, \theta_s, \theta_t, s) & = - \dfrac{1}{240 (\nu + 2\frac{\theta_s - 2\theta_t}{3})^2}
- \dfrac{1}{240 (\nu + 2\frac{\theta_t - 2\theta_s}{3})^2}
- \dfrac{1}{240 (\nu + 2\frac{\theta_s + \theta_t}{3})^2} + O(s^{-3}), 
\label{PIVlogexp2}
\end{split}
\end{equation}
and the higher genus contributions can be computed analogously.


\section{Magnetic and dyonic expansions of Argyres-Douglas prepotentials} 
\label{gauge}

In this section we consider the four-dimensional $\CN=2$ theories associated to the previously discussed PI, PII and PIV equations, i.e. Argyres-Douglas theories $H_0$, $H_1$ and $H_2$, and compute the lowest genera prepotentials $\mathcal{F}_g$ at the various strongly coupled regimes. 
These ``magnetic" or ``dyonic" expansions will be identified with to the long-distance expansions of the corresponding Painlev\'{e} $\tau$-function. The same analysis for $SU(2)$ SQCD will be performed in appendix \ref{AppB}.
  
We consider the partition function of a rather special Omega background $\epsilon_1 = - \epsilon_2 \equiv \epsilon$ where the connection with topological string theory is natural.
The partition function of the gauge theory in this Omega background can be written as
    \bea
    Z
     =     \exp \left(\sum_{g=0}^\infty  \epsilon^{2g-2} \CF_g \right).
    \eea
We will refer to $\CF_g$ as genus $g$ prepotential in the following.
  
The genus zero prepotential determines the low energy effective theory of the $\CN=2$ theory where the nontrivial background is turned off.
This is determined by the Seiberg-Witten curve, which is defined on Coulomb branch parametrized by $u$, and the differential $\lambda(u)$ on it \cite{Seiberg:1994rs}:
    \bea
    a_i (u)
     =     \oint_{A_i} \lambda(u), ~~~~
    \frac{\partial \CF_0}{ \partial a_i} (u)
     =     \oint_{B_i} \lambda(u).
    \eea
    
In particular the Seiberg-Witten curve for $\mathcal{N} = 2$ $SU(2)$ SQCD with $N_f = 0, \ldots, 3$ can be written in the general form
\begin{equation}
y^2 = x^4 + A x^3 + B x^2 + C x + D = \prod_{i=1}^4 (x - e_i),
\end{equation}
with $e_i$ zeroes of the curve; the curves for Argyres-Douglas theories can be obtained from the SQCD ones by taking appropriate scaling limits. From the curve (and some additional imput) we can extract not only the genus zero (\cite{Seiberg:1994rs} for $N_f = 0$, \cite{1997IJMPA..12.3413M} for $N_f = 1,2,3$) but also higher genus (\cite{2007JHEP...09..054H} for $N_f = 0$, \cite{2010JHEP...07..083H} for $N_f = 1,2,3$) prepotentials in the strong-coupling regimes. In order to do that we need to introduce the combination
\begin{equation}
\Delta = \prod_{i<j}(e_i-e_j)^2 ,
\end{equation}
which can be alternatively written as
\begin{equation}
\begin{split}
\Delta = & -27A^4D^2 + A^3C(18BD - 4C^2) + AC(18BC^2 - 80B^2D - 192D^2) \\
& + A^2(B^2C^2 + 144BD^2 - 4B^3D - 6C^2D) - 4B^3C^2 - 27C^4 \\
& + 16B^4D + 144BC^2D - 128B^2D^2 + 256D^3. \label{Delta}
\end{split}
\end{equation}
In addition, we define
\begin{equation}
\mathcal{D} = \dfrac{1}{2}\sum_{i<j}(e_i - e_j)^2 = -B^2 + 3AC - 12D.
\end{equation}
In the strong coupling regime ($a, a_D$ have different meaning according to the expansion point considered: magnetic, dyonic, \ldots) we have \cite{1997IJMPA..12.3413M}
\begin{eqnarray}
\dfrac{d a}{d u} &=& \dfrac{\sqrt{2}}{2}(-\mathcal{D})^{-1/4} \left[ \dfrac{3}{2\pi} \ln 12 
\;{}_{2}F_{1} \left( \dfrac{1}{12}, \dfrac{5}{12}; 1; -\dfrac{27 \Delta}{4 \mathcal{D}^3} \right) 
- \dfrac{1}{2\pi} F^* \left( \dfrac{1}{12}, \dfrac{5}{12}; 1; -\dfrac{27 \Delta}{4 \mathcal{D}^3} \right) \right], \nonumber \\
\dfrac{d a_D}{d u} &=& i \dfrac{\sqrt{2}}{2}(-\mathcal{D})^{-1/4} {}_{2}F_{1} \left( \dfrac{1}{12}, \dfrac{5}{12}; 1; -\dfrac{27 \Delta}{4 \mathcal{D}^3} \right) ,\label{F0}
\end{eqnarray}
with
\begin{equation}
F^*(\alpha, \beta; 1; z) = {}_{2}F_{1} \left( \alpha, \beta; 1; z \right ) \ln z + 
\sum_{n=0}^{\infty} \dfrac{(\alpha)_n(\beta)_n}{(n!)^2} \sum_{k=0}^{n-1} \left[ \dfrac{1}{\alpha + n} + \dfrac{1}{\beta + n} - \dfrac{2}{n + 1} \right].
\end{equation}
By integrating these series with respect to $u$ and expressing everything in terms of $a_D$, one can easily obtain the genus zero prepotential $\mathcal{F}_0$. For the genus one prepotential $\mathcal{F}_1$ we can use the formula
\begin{equation}
\mathcal{F}_1 = - \dfrac{1}{12} \ln \left(\dfrac{d a_D}{d u}\right) - \dfrac{1}{12} \ln \Delta, \label{F1}
\end{equation}
obtained in \cite{2007JHEP...09..054H}, \cite{2010JHEP...07..083H}; in the same papers we can find the higher genus prepotentials $\mathcal{F}_g$, $g \geqslant 2$ (although only for the massless case).



\subsection{Argyres-Douglas theory $H_0$}

Let us start from the $H_0$ theory thought as coming from the conformal point of $SU(2)$ gauge theory with $N_f=1$; as discussed in \cite{Argyres:1995xn}, this fixed point is the same as the one studied in \cite{Argyres:1995jj}.
The Seiberg-Witten curve is given by 
\begin{equation}
y^2 = z^3 - 3 c z + u,
\end{equation} 
and the Seiberg-Witten differential is $\lambda = y dz$. This curve can be easily seen to coincide with \eqref{trA2PI} which was obtained from the PI Lax pair, after an appropriate rescaling and identification of the various terms.  
The parameters $u$ and $c$ in the curve are the vev of the relevant operator $\CO$ from the conformal point
and the corresponding coupling $\int d^2 \theta_1 d^2 \theta_2 c \CO $, and have dimensions $6/5$ and $4/5$ respectively. Therefore $u$ parametrizes the Coulomb branch of the theory.
The discriminant of the curve is given by 
\begin{equation}
\Delta = 27(4c^3 - u^2).
\end{equation}
Its zeroes tell us the position of the singularities on the Coulomb branch moduli space; they are located at
\begin{equation}
u_1 = -2c^{3/2} ,\;\;\; u_2 = 2c^{3/2}.
\end{equation}
In the limit $c \rightarrow 0$, the singularities collide to give the conformal point.
Instead we are interested in the other limit where $c$ is large and consider the expansion around $u_1$ or $u_2$. \\

\noindent \textbf{$H_0$ expansion 1} \\

\noindent The genus zero prepotential around $u_1$ can be easily computed and it is given by 
\begin{equation}
\begin{split}
2\pi i \mathcal{F}_0 \;=\; & \frac{a_D^2}{4} \log \left(-\frac{a_D^2}{248832 c^{5/2}}\right) - \frac{3 a_D^2}{4}
+ b_1 a_D c^{5/4} + b_2 c^{5/2} \\
& - \frac{47 a_D^3}{288 \sqrt{3} c^{5/4}}+\frac{7717 a_D^4}{497664 c^{5/2}} + O\left(a_D^5\right),
\end{split}
\end{equation}  
with $b_1$, $b_2$ integration constants. The genus one prepotential is 
\begin{equation}
\mathcal{F}_1 =  - \frac{1}{12} \ln \dfrac{a_D}{6 \sqrt{3} i c^{5/4}} -\frac{17 a_D}{576 \sqrt{3} c^{5/4}}+ \frac{3677 a_D^2}{497664 c^{5/2}} + O(a_D^3).
\end{equation}
These coincide with the PI results of section \ref{SecPI} after identifying the parameters as $\nu = a_D$ and $s = 6 \sqrt{3} i \, c^{5/4}$. \\

\noindent \textbf{$H_0$ expansion 2} \\

\noindent The genus zero prepotential around $u_2$ is given by 
\begin{equation}
\begin{split}
2\pi i \mathcal{F}_0 \;=\; & \frac{a_D^2}{4} \log \left(-\frac{a_D^2}{248832 c^{5/2}}\right) - \frac{3 a_D^2}{4} 
+ b_1' a_D c^{5/4} + b_2' c^{5/2} \\
& + \frac{47 a_D^3}{288 \sqrt{3} c^{5/4}}+\frac{7717 a_D^4}{497664 c^{5/2}} + O\left(a_D^5\right).
\end{split}
\end{equation}  
The genus one prepotential is 
\begin{equation}
\mathcal{F}_1 =  - \frac{1}{12} \ln \dfrac{-a_D}{6 \sqrt{3} i c^{5/4}} +\frac{17 a_D}{576 \sqrt{3} c^{5/4}}+ \frac{3677 a_D^2}{497664 c^{5/2}} + O(a_D^3).
\end{equation}
These are the same expressions as above, where $\frac{a_D}{c^{5/4}} \rightarrow - \frac{a_D}{c^{5/4}}$.
    
\subsection{Argyres-Douglas theory $H_1$}

There are two possible realizations for the Seiberg-Witten curve of $H_1$; the first one is
\begin{equation}
y^2 = z^4 + 4 c z^2 + 2 m z + u .\label{firstII}
\end{equation}
It coincides with the expression \eqref{trA2PII} obtained from the Jimbo-Miwa PII Lax pair. The second realization reads
\begin{equation}
y^2 = z + c + \frac{u}{z} + \frac{m^2}{z^2},
\end{equation}
and can be matched with \eqref{trA2PIIFN} corresponding to the Flaschka-Newell PII Lax pair. 
The $H_1$ theory has an $SU(2)$ flavor symmetry whose mass parameer is $m$.
The role of $u$ and $c$ are the same as that of the $H_0$ theory.
We will focus on the first realization \eqref{firstII} in the following. Its discriminant
\begin{equation}
\Delta = 256 u^3 - 2048 u^2 c^2 + 4096 c^4 u + 2304 m^2 c u - 432 m^4 - 1024 m^2 c^3
\end{equation}
vanishes at (perturbatively in $m$ small)
\begin{equation}
\begin{split}
u_1 &= \dfrac{m^2}{4c} + O(m^4), \\
u_2 &= 4c^2 + 2\sqrt{2}i m c^{1/2} - \dfrac{m^2}{8 c} + \dfrac{i m^3}{64 \sqrt{2} c^{5/2}} + O(m^4), \\
u_3 &= 4c^2 - 2\sqrt{2}i m c^{1/2} - \dfrac{m^2}{8 c} - \dfrac{i m^3}{64 \sqrt{2} c^{5/2}} + O(m^4).
\end{split}
\end{equation}
We therefore have three expansions, with the second and third one related by a $\mathbb{Z}_2$ symmetry $m \leftrightarrow -m$. \\

\noindent \textbf{$H_1$ expansion 1} \\

\noindent The genus zero prepotential around $u_1$ is given by 
\begin{equation}
\begin{split}
2\pi i \mathcal{F}_0 \;=\; & \frac{a_D^2}{4} \ln \left(-\frac{8192 c^{3}}{a_D^2}\right) + \frac{3 a_D^2}{4} 
+ b_1 a_D c^{3/2} + b_2 c^3 \\
& + \frac{17 i a_D^3}{192 \sqrt{2} c^{3/2}} + \frac{125 a_D^4}{16384 c^{3}} + \dfrac{13 a_D^2 m^2}{128 c^3} + O\left(a_D^5\right).
\end{split}
\end{equation}
The genus one prepotential reads
\begin{equation}
\mathcal{F}_1 = -\dfrac{1}{12} \ln \dfrac{-3 i a_D}{16 \sqrt{2} c^{3/2}} - \frac{31 i a_D}{384 \sqrt{2} c^{3/2}} 
- \frac{293 a_D^2}{16384 c^{3}} - \dfrac{5 m^2}{768 c^3} + O\left(a_D^3\right).
\end{equation}
These agree with the PII first expansion of section \ref{SecPII} for $\nu = - i a_D$, $s = \frac{16 \sqrt{2}}{3}c^{3/2}$, $\theta = \sqrt{2} m$. \\

\noindent \textbf{$H_1$ expansion 2,3} \\

\noindent The genus zero prepotential around $u_2 = u_3$ is quite hard to compute in gauge theory for generic $m$, so we will just present the result for $m=0$ which is given by 
\begin{equation}
\begin{split}
2\pi i \mathcal{F}_0 \;=\; & -\dfrac{a_D^2}{2} \ln \left( - \dfrac{a_D^2}{4096 c^3} \right) + \frac{3 a_D^2}{2} 
+ b_1' a_D c^{3/2} + b_2' c^3 \\
& - \frac{17 i a_D^3}{96 c^{3/2}} + \frac{125 a_D^4}{4096 c^{3}} + O\left(a_D^5\right).
\end{split}
\end{equation}
On the other hand, it is possible to compute the genus one prepotential at generic $m$; its expansion around $u_2$ reads
\begin{equation}
\mathcal{F}_1 = -\dfrac{1}{12} \ln \left(3\dfrac{\tilde{a}_D + \frac{m}{\sqrt{2}}}{32c^{3/2}} \right)
-\dfrac{1}{12} \ln \left( 3\dfrac{\tilde{a}_D - \frac{m}{\sqrt{2}}}{32c^{3/2}} \right) + \frac{i \tilde{a}_D}{192 c^{3/2}} 
- \frac{15 \tilde{a}_D^2}{4096 c^{3}} + \dfrac{17 m^2}{24576 c^3} + O\left(\tilde{a}_D^3\right),
\end{equation}
with
\begin{equation}
\tilde{a}_D = a_D - \dfrac{m}{\sqrt{2}},
\end{equation}
while the one around $u_3$ is given by the same formula with
$\tilde{a}_D = a_D + \dfrac{m}{\sqrt{2}}$.
These  formulae agree with the  second PII tau function expansion of section \ref{SecPII} under parameter identification $\nu = - i \tilde{a}_D$, $s = \frac{32 i}{3}c^{3/2}$, $\theta = \pm i \sqrt{2} m$.

\subsection{Argyres-Douglas theory $H_2$}

The Seiberg-Witten curve reads
\begin{equation}
y^2 = z^2 + 2cz + (2\tilde{m} + c^2) + \dfrac{u + 2c m_-}{z} + \dfrac{m_-^2}{z^2},
\end{equation}
where $m_-$ and $\tilde{m}$ are the mass parameters of the $SU(3)$ flavor symmetry.
The zeroes of the discriminant \eqref{Delta} with
\begin{equation}
\begin{split}
A & \,=\, 2 c, \\
B & \,=\, c^2 + 2 \tilde{m}, \\
C & \,=\, u + 2 c \tilde{m}, \\
D & \,=\, m_-^2 ,
\end{split}
\end{equation}
are located at (perturbatively in $\tilde{m}$, $m_-$ small)
\begin{equation}
\begin{split}
u_1 & = \dfrac{4c^3}{27} - \dfrac{4 c \tilde{m}}{3} + \dfrac{3 m_-^2 + \tilde{m}^2}{c} + O(c^{-3}), \\
u_2 & = - 2 c(\tilde{m} + m_-) - \dfrac{2 m_-(\tilde{m} + m_-)}{c} + O(c^{-3}), \\
u_3 & = \dfrac{m_-^2 - \tilde{m}^2}{c} + O(c^{-3}), \\
u_4 & = - 2 c(\tilde{m} - m_-) + \dfrac{2m_-(\tilde{m} - m_-)}{c} + O(c^{-3}). \\
\end{split}
\end{equation}
We therefore expect four expansions for generic values of the masses, with the first one qualitatively different from the other three (judging from the massless limit). Since computations with all masses turned on are too involved, here we present the results for $\mathcal{F}_0$ and $\mathcal{F}_1$ at zero masses, case in which $u_2 = u_3 = u_4$ and we only have two expansions. \\

\noindent \textbf{$H_2$ expansion 1 (massless)} \\

\noindent The genus zero prepotential around $u_1$ reads
\begin{equation}
\begin{split}
2\pi i \mathcal{F}_0 \;=\; & - \dfrac{a_D^2}{4} \ln \left( -\dfrac{a_D^2}{24 c^4} \right)
+ \dfrac{3}{4} a_D^2 + b_1 a_D c^2 + b_2 c^4 \\
& - \dfrac{i a_D^3}{\sqrt{6} c^2} + \dfrac{35a_D^4}{96 c^4} + O\left(a_D^5\right).
\end{split}
\end{equation}
The genus one prepotential instead is
\begin{equation}
\mathcal{F}_1 = -\dfrac{1}{12} \ln \left( - \dfrac{\sqrt{3}a_D}{\sqrt{2}c^2} \right) 
+ \frac{i \sqrt{2} a_D}{\sqrt{3} c^{2}} - \dfrac{181 a_D^2}{96 c^4} + O\left(a_D^3\right).
\end{equation}
These expressions coincide with the results of the first PIV tau function expansion in section \ref{SecPIV} for $\nu = i a_D$, $s = -i \sqrt{\frac{2}{3}}c^2$. \\

\noindent \textbf{$H_2$ expansion 2,3,4 (massless)} \\

\noindent The genus zero prepotential around $u_2 = u_3 = u_4$ (with integration constants) can be easily computed and it is given by 
\begin{equation}
\begin{split}
2\pi i \mathcal{F}_0 \;=\; & - \dfrac{3 a_D^2}{4} \ln \left( -\dfrac{a_D^2}{8 c^4} \right)
+ \dfrac{9}{4} a_D^2 + b_1' a_D c^2 + b_2' c^4 \\
& + \dfrac{3 i a_D^3}{\sqrt{2} c^2} + \dfrac{105a_D^4}{32 c^4} + O\left(a_D^5\right).
\end{split}
\end{equation}
The genus one prepotential reads
\begin{equation}
\mathcal{F}_1 = -\dfrac{1}{4} \ln \left( -\dfrac{i a_D}{\sqrt{2}c^2} \right)  
- \frac{3 a_D^2}{32 c^{4}} + O\left(a_D^3\right).
\end{equation}
These expressions coincide with the results of the second expansion in section \ref{SecPIV} for $\nu = i a_D$, $s = - \sqrt{2}c^2$. \\


\section{Conclusions and discussions} 
\label{concl}
 We have studied in this paper the correspondence between four-dimensional $\mathcal{N} = 2$ rank-one theories and Painlev\'{e} transcendents. As argued in section \ref{sw}, there is a direct relationship between them bridged by a Riemann sphere with punctures and the Hitchin system on it. 
  In particular, we found that the Painlev\'{e} IV, II and I correspond 
  to the rank-one SCFTs of Argyres-Douglas type, so-called $H_2$, $H_1$ and $H_0$ respectively.
  The long-distance expansions of Painlev\'{e} tau functions have been checked to agree 
  with the magnetic or dyonic expansions of the partition functions of the corresponding four-dimensional theories.
  
  While we have seen the agreement by expanding both functions, the correspondence is not completely clear. 
  Here we discuss some interesting directions to show the correspondence, and generally to related topics:
\begin{itemize}
    \item There is an important difference between the four-dimensional theory side and the Painlev\'{e} side,
             which is the $\rho$ variable in \eqref{fourierNf4}.
             This does not usually appear in the four-dimensional theory side, 
             rather in the $\rho$ variable 
             with the B-period $a_D$, 
             as in \eqref{fourierNf4}.
             It would be interesting to understand better the four-dimensional theory origin of this additional parameter.
   \item Although somehow clear from the point of view of the integrable system associated to the Higgs 
             bundle/Hitchin system, we are still lacking a good gauge theory understanding of the monodromy
             preserving condition for a deformation of the oper connection.
   \item In section \ref{sw} we discussed the theory of isomonodromic deformation with connection $\mathbf{A}$ 
             and the Hitchin moduli space, the space of the flat connection $\mathcal{A}$.
             The sections of $\mathbf{A}$, that is the solutions \eqref{solCFT} to the system of linear ODE,
             will now be those of the flat connection $\mathcal{A}$ in the oper limit.
             The canonical sections of $\mathcal{A}$ and of its oper limit can be used to construct 
             triangulations of the punctured sphere $\mathcal{C}_{0,n}$ on which the connection lives; this 
             triangulation might be the WKB or the Fock-Goncharov one \cite{2009arXiv0907.3987G}, although in 
             \cite{1511.03851} the authors explicitly affirmed that their triangulation is different
             from the one in \cite{2009arXiv0907.3987G}. It would be interesting to understand better which
             triangulation is the most natural one in the context of Painlev\'{e} theory; it might also 
             be interesting to see if one can give an explicit expression for these sections similar to what
             done here for the tau function.               
    \item The differential Painlev\'{e} equations we studied in this paper are a subset of a bigger family of
             equations which also involves the finite-difference, $q$-difference and elliptic Painlev\'{e} 
             equations. All of them are classified in \cite{Sakai2001} based on rational surfaces and
             associated actions of affine Weyl groups as in Table \ref{tab:sakai}; the corresponding
             classification by symmetries is listed in Table \ref{tab:sakaisym}. 
             More precisely the ones in the box, corresponding to the confluence diagram in Figure
             \ref{coalescence}, are the differential Painlev\'{e} equations we considered, while the one on
             the left of the box are the difference Painlev\'{e} \cite{2007arXiv0706.2634B}; above the box we 
             have instead the $q$-difference Painlev\'{e} as well as the only elliptic one (the upper left 
             entry).
             
             \renewcommand{\arraystretch}{1.3}
             \begin{table}[t]
             \centering
             \scalebox{0.8}[0.8]{
             \begin{tabular}{ccccccccccccccccc}
	    $A_0^{(1)}$ &&&&&&&&&&&&&& $A_7^{(1)}$ &&\\
	    $\downarrow$ &&&&&&&&&&&&& $\nearrow$ && $\searrow$ &\\
	    $A_0^{(1)}$ & $\rightarrow$ & $A_1^{(1)}$ & $\rightarrow$ & $A_2^{(1)}$ & $\rightarrow$ & $A_3^{(1)}$ & $\rightarrow$ & $A_4^{(1)}$ & $\rightarrow$ & $A_5^{(1)}$ & $\rightarrow$ & $A_6^{(1)}$ & $\rightarrow$ & $A_7^{(1)'}$ & & $A_8^{(1)}$ \\
	    &&&&&&& $\searrow$ && $\searrow$ && $\searrow$ && $\searrow$ && $\searrow$ &\\ \cline{9-17}
	    && $A_0^{(1)}$ & $\rightarrow$ & $A_1^{(1)}$ & $\rightarrow$ & $A_2^{(1)}$ & $\rightarrow$ & \multicolumn{1}{|c}{$D_4^{(1)}$} & $\rightarrow$ & $D_5^{(1)}$ & $\rightarrow$ & $D_6^{(1)}$ & $\rightarrow$ & $D_7^{(1)}$ & $\rightarrow$ & \multicolumn{1}{c|}{$D_8^{(1)}$} \\
	    &&&&&&&&\multicolumn{1}{|c}{}&&& $\searrow$ && $\searrow$ &&$\searrow$& \multicolumn{1}{c|}{} \\
	    &&&&&&&&\multicolumn{1}{|c}{}&&&& $E_6^{(1)}$ & $\rightarrow$ & $E_7^{(1)}$ & $\rightarrow$ & \multicolumn{1}{c|}{$E_8^{(1)}$} \\ \cline{9-17}
	    \end{tabular}}
	    \caption{The list of affine root systems associated to discrete Painlev\'{e} equations.} 
	    \label{tab:sakai}
	    \end{table}
	    \renewcommand{\arraystretch}{1}
	    \renewcommand{\arraystretch}{1.3}
	    \begin{table}[t]
	    \small
             \centering
             \scalebox{0.8}[0.8]{
             \begin{tabular}{ccccccccccccccccc}
	    $E_8^{(1)}$ &&&&&&&&&&&&&& $A_1^{(1)}$ &&\\
	    $\downarrow$ &&&&&&&&&&&&& $\nearrow$ && $\searrow$ &\\
	    $E_8^{(1)}$ & $\rightarrow$ & $E_7^{(1)}$ & $\rightarrow$ & $E_6^{(1)}$ & $\rightarrow$ & $D_5^{(1)}$ & $\rightarrow$ & $A_4^{(1)}$ & $\rightarrow$ & $(A_1+A_2)^{(1)}$ & $\rightarrow$ & $(A_1+A_1)^{(1)}$ & $\rightarrow$ & $A_1^{(1)}$ & & $A_0^{(1)}$ \\
	    &&&&&&& $\searrow$ && $\searrow$ && $\searrow$ && $\searrow$ && $\searrow$ &\\ \cline{9-17}
	    && $E_8^{(1)}$ & $\rightarrow$ & $E_7^{(1)}$ & $\rightarrow$ & $E_6^{(1)}$ & $\rightarrow$ & \multicolumn{1}{|c}{ $D_4^{(1)}$ } & $\rightarrow$ & $A_3^{(1)}$ & $\rightarrow$ & $(A_1+A_1)^{(1)}$ & $\rightarrow$ & $A_1^{(1)}$ & $\rightarrow$ & \multicolumn{1}{c|}{$A_0^{(1)}$} \\ 
	    &&&&&&&&\multicolumn{1}{|c}{}&&& $\searrow$ && $\searrow$ & &$\searrow$& \multicolumn{1}{c|}{} \\
	    &&&&&&&&\multicolumn{1}{|c}{}&&&& $A_2^{(1)}$ & $\rightarrow$ & $A_1^{(1)}$ & $\rightarrow$ & \multicolumn{1}{c|}{$A_0^{(1)}$} \\ \cline{9-17}
	    \end{tabular}}
	    \caption{The list of the symmetries of discrete Painlev\'{e} equations.} 
	    \label{tab:sakaisym}
	    \end{table}
	    \renewcommand{\arraystretch}{1}
    	    
	    For the differential Painlev\'{e}, the symmetry is the same as (the non-Affine version of) the flavor
	    symmetry of the corresponding four-dimensional theory.
	    In a similar manner, one would notice that (the non-affine version of) the symmetry of the
	    finite-difference equations at the left of the box is identified with the flavor symmetry of the 
	    four-dimensional Minahan-Nemeschansky theories \cite{1996NuPhB.482..142M,1997NuPhB.489...24M}. 
	    Moreover, the symmetry of the $q$-difference Painlev\'{e} associated to $A_f^{(1)}$ above the box in 
	    Table \ref{tab:sakai} is identified with the flavor symmetry of the five-dimensional $SU(2)$ theory 
	    with $8-f$ flavors \cite{1996PhLB..388..753S} (with two possibilities of theta-angle $\theta = 0, \pi$ 
	    for the $f=0$ theory) as well as ``local $\mathbb{F}_0$'' \cite{Kim:2014nqa}; the upmost left one, 
	    that is the only elliptic equation, would then correspond to the five-dimensional $SU(2)$ theory with 
	    $N_F = 8$ which is actually believed to have a UV completion in six dimensions. Therefore, the five-dimensional 
	    partition functions of these theories are expected to be related to the solutions of the discrete Painlev\'{e} equation.
	    Indeed in \cite{2016arXiv160802566B,BGTII} it was argued that the latter agrees with $q$-deformed Virasoro 
	    conformal blocks. With help of the five-dimensional AGT correspondence \cite{Awata:2009ur} 
	    this is related to the Nekrasov partition function of five-dimensional $SU(2)$ theory.
	    It would be interesting to pursue this point further.                           

    \item At the level of gauge theory it is very natural to have two parameters $\epsilon_1$, $\epsilon_2$; 
        translated in CFT language this corresponds to moving away from $c = 1$. The 
        combination $\epsilon_1 + \epsilon_2$ is expected to arise as the Planck constant when promoting $q$, $p$ 
        and the Hamiltonian $\mathcal{H}(t)$ to quantum operators; this however takes us away from the theory of 
        Painlev\'{e} equations and might involve some quantum version of them.  
        It would be interesting to understand if the corresponding dynamics can be formulated
as an isomonodromic problem for a corresponding quantum Hitchin system and in such a context
give an interpretation to the quantum $\hat{p}$ and $\hat{q}$ operators.    
    \item Isomonodromic deformation problems can be formulated on punctured Riemann surfaces of any genus and 
        they can involve connections on any rank; it would therefore be natural to study these more general 
        problems in terms of CFT or four-dimensional quiver theories along the lines 
        pursued in this paper, taking into account the fact that in this case there will be more deformation 
        parameters. 
        An initial study of higher rank connections on a punctured Riemann sphere appeared in 
        \cite{2015JHEP...09..167G}; the higher genus case seems to be much less studied in the literature, 
        apart from the genus one case with one regular puncture.
    \item The magnetic phase of the pure SU(2) SYM theory can be described in terms of a
        Fermi gas and a resulting matrix model \cite{Bonelli:2016idi}. It would be interesting to investigate 
        how to use the results of this paper to extend to other examples the Fermi gas approach.
    \item relation to BPS spectra counting and quivers via triangulation of (bordered) Riemann surface
    \item a proposal for $S^4$ partition function of Argyres-Douglas theories can be formulated by using the 
        results on Painlev\'{e}. These provide a natural candidate for the perturbative part of gauge theory 
        partition function around AD points and we can use it to make a proposal for the squashed $S^4$ (recall 
        that for Painlev\'{e} $\epsilon_1+\epsilon_2=0$) partition function in this regime.

\end{itemize}

\section*{Acknowledgements}

We would like to thank Misha Bershtein, Sergio Cecotti, Bernard Julia, Piljin Yi for fruitful and insightful discussions.
K.~M.~and A.~T.~would like to thank the theory group in \'{E}cole Normale Sup\'{e}rieure for warm hospitality.
K.~M.~would like to thank ICTP and SISSA for kind hospitality during the course of the project.
A.S. would like to thank the Perimeter Institute for its very kind hospitality during the course of this project. 
This research was supported in part by Perimeter Institute for Theoretical Physics. Research at Perimeter Institute is supported by the Government of Canada through Industry Canada and by the Province of Ontario through the Ministry of Research and Innovation.
The work of G.B. is supported by the INFN Iniziativa Specifica ST\&FI.
The work of A.T. is supported by the INFN Iniziativa Specifica GAST.

\appendix

\section{Appendix A: Long-distance expansions for PIII$_{1,2,3}$ and PV} 
\label{AppA}

In this Appendix we collect the long-distance expansions for the $\tau$-functions of PIII$_3$, PIII$_2$, PIII$_1$ and PV which were not considered in Section \ref{Painleve}.

\subsection{Painlev\'{e} III$_{3}$} \label{SecPIII3}

A convenient Lax pair for PIII$_{3}$ can be obtained by slightly modifying the one given in \cite{2009arXiv0902.1702V} and is given by
\begin{equation}
{\bf A} = A_0 + \dfrac{A_1}{z}  + \dfrac{A_2}{z^2}  =
\left(\begin{array}{cc}
\frac{pq}{z} & 1-\frac{t}{zq} \\ 
\frac{1}{z} - \frac{q}{z^2} & -\frac{pq}{z} 
\end{array} \right) ,
\end{equation}
\begin{equation}
{\bf B} = B_0 + \dfrac{B_1}{z}  =
\left(\begin{array}{cc}
0 & \;\; \frac{1}{q} \\ 
\frac{q}{t z} & \;\; 0 
\end{array} \right) .
\end{equation}
The compatibility condition \eqref{cc} requires
\renewcommand\arraystretch{1.8}
\begin{equation}
\left\{
\begin{array}{l}
\dot{q} = -\dfrac{2pq^2}{t} + \dfrac{q}{t}, \\
\dot{p} = \dfrac{2pq^2}{t} - \dfrac{p}{t} + \dfrac{1}{q^2} - \dfrac{1}{t} ,
\end{array} \right. \label{qpIII3}
\end{equation}
\renewcommand\arraystretch{1}
and leads to the PIII$_{3}$ equation
\begin{equation}
\ddot{q} = \dfrac{\dot{q}^2}{q} - \dfrac{\dot{q}}{t} + \dfrac{2q^2}{t^2} - \dfrac{2}{t}. \label{PIII3}
\end{equation}
We can now take the trace 
\begin{equation}
\dfrac{1}{2} \operatorname{Tr}{\bf A}^2 = \dfrac{t}{z^3} + \dfrac{\sigma_{III_3}(t)}{z^2} + \dfrac{1}{z}, \label{trA2PIII3}
\end{equation}
where
\begin{equation}
\sigma_{III_3}(t) = p^2q^2 - q - \dfrac{t}{q}.
\end{equation}
The function $\sigma_{III_3}(t) = t\frac{d}{dt} \ln \tau_{III_3}(t)$ satisfies the $\sigma$-PIII$_{3}$ Painlev\'{e} equation
\begin{equation}
(t\ddot{\sigma}_{III_3})^2 = 4 (\dot{\sigma}_{III_3})^2 \left( \sigma_{III_3} - t \dot{\sigma}_{III_3} \right) - 4 \dot{\sigma}_{III_3} \label{eqsigmaPIII3}
\end{equation}
with respect to the dynamics given by \eqref{qpIII3}. 
We find the following expansions for $\tau_{III_3}(t)$ as 
$t\to +\infty$, cf \cite[Eqs. (3.13)-(3.15)]{2014arXiv1403.1235I}: \\

\noindent \textbf{$\tau$-PIII$_{3}$ expansion} \\

\noindent The asymptotic expansion of the tau function on the ray arg$\,t = 0$ (i.e. $s \in \mathbb{R}$) reads
\begin{equation}
\begin{split}
& \tau_{III_3}(t) = s^{\frac{1}{6}} \sum_{n \in \mathbb{Z}} e^{i n \rho} \mathcal{G}(\nu + n, s), \;\;\;\;\;\; t = 2^{-12}s^4, \\
& \mathcal{G}(\nu, s) = C(\nu,s) \left[ 1 + \sum_{k=1}^{\infty} \dfrac{D_k(\nu)}{s^k} \right], \\
& C(\nu,s) = (2\pi)^{-\frac{\nu}{2}} e^{\frac{s^2}{16} + i \nu s + \frac{i \pi \nu^2}{4}} s^{\frac{1}{12} - \frac{\nu^2}{2}} 2^{-\nu^2} G(1+\nu), \label{III31}
\end{split}
\end{equation} 
where the first few coefficients are given by
\begin{equation}
\begin{split}
& D_1(\nu) = - \dfrac{i \nu (2 \nu^2 - 1)}{8}, \\
& D_2(\nu) = - \dfrac{\nu^2 (4\nu^4 + 16 \nu^2 - 11)}{128}.
\end{split}
\end{equation}
From the number of Barnes functions we see that there is only one light particle in this sector. We therefore get
\begin{equation}
\ln \left[ \mathcal{G}\left(\frac{\nu}{\epsilon}, \frac{s}{\epsilon}\right) \right] \;=\;
\sum_{g \geqslant 0} \epsilon^{2g-2} \mathcal{F}_g(\nu, s) 
\end{equation}
with 
\begin{equation}
\begin{split}
\mathcal{F}_0(\nu, s) & = \dfrac{s^2}{16} + i \nu s + \dfrac{\nu^2}{2} \ln \dfrac{i \nu}{4 s} 
- \dfrac{3\nu^2}{4} - \dfrac{i \nu^3}{4 s} - \dfrac{5 \nu^4}{32 s^2} + O(s^{-3}), \\
\mathcal{F}_1(\nu, s) & = \zeta'(-1) - \dfrac{1}{12}\ln \dfrac{\nu}{s} + \dfrac{i \nu}{8 s} + \dfrac{3 \nu^2}{32 s^2} + O(s^{-3}), \\
\mathcal{F}_2(\nu, s) & = - \dfrac{1}{240 \nu^2} + O(s^{-3}), \\
\mathcal{F}_g(\nu, s) & = \ldots. 
\label{PIII3logexp1}
\end{split}
\end{equation}


\subsection{Painlev\'{e} III$_{2}$} \label{SecPIII2}

Again, we will take as the Lax pair for PIII$_{2}$ the one obtained by slightly modifying the Lax pair given in \cite{2009arXiv0902.1702V}; explicitly, we have
\begin{equation}
\begin{split}
& {\bf A} = A_0 + \dfrac{A_1}{z}  + \dfrac{A_2}{z^2}  = \\
& = \left(\begin{array}{cc}
\frac{t}{2} + \frac{\theta_*}{z} + \frac{2 p q^2 - 2 q\theta_* - t q^2}{2z^2} & \;\;\;
\frac{4p^2 q^4 - 4tpq^4 +t^2 q^4 -4q^2 \theta_*^2 - 4q}{4 q^2 z} + \frac{\left( 2pq^2 - tq^2 - 2q\theta_* \right)^2}{4 q z^2} \\ 
\frac{1}{z} - \frac{q}{z^2} & - \frac{t}{2} - \frac{\theta_*}{z} - \frac{2 p q^2 - 2 q\theta_* - t q^2}{2z^2} 
\end{array} \right),
\end{split} 
\end{equation}
\begin{equation}
{\bf B} = B_0 + B_1 z=
\left(\begin{array}{cc}
\frac{z}{2} + \frac{tq + 2\theta_*}{2t} & 
\;\; \frac{4p^2q^4 -4tpq^4 + t^2q^4 - 4q^2 \theta_*^2 - 4q}{4tq^2} \\ 
\frac{1}{t} & \;\; -\frac{z}{2} - \frac{tq + 2\theta_*}{2t} 
\end{array} \right) .
\end{equation}
The compatibility condition \eqref{cc} requires
\renewcommand\arraystretch{1.8}
\begin{equation}
\left\{
\begin{array}{l}
\dot{q} = -\dfrac{2pq^2}{t} ,\\
\dot{p} = \dfrac{2pq^2}{t} + \dfrac{1}{tq^2} - \dfrac{t q}{2} + \dfrac{1}{2} - \theta_* ,
\end{array} \right. \label{qpIII2}
\end{equation}
\renewcommand\arraystretch{1}
and leads to the PIII$_{2}$ equation
\begin{equation}
\ddot{q} = \dfrac{\dot{q}^2}{q} - \dfrac{\dot{q}}{t} + q^3 - \dfrac{q^2(1-2\theta_*)}{t} - \dfrac{2}{t^2}. \label{PIII2}
\end{equation}
The trace 
\begin{equation}
\dfrac{1}{2} \operatorname{Tr}{\bf A}^2 = \dfrac{1}{z^3} + \dfrac{\sigma_{III_2}}{z^2} + \dfrac{t \theta_*}{z} + \dfrac{t^2}{4} \label{trA2PIII2}
\end{equation}
contains the function
\begin{equation}
\sigma_{III_2}(t) = p^2q^2 - \frac{q^2 t^2}{4} - t q \theta_* - \dfrac{1}{q},
\end{equation}
whih satisfies the $\sigma$-PIII$_{2}$  equation
\begin{equation}
(t\ddot{\sigma}_{III_2})^2 = 4 (\dot{\sigma}_{III_2})^2 \left( \sigma_{III_2} - t \dot{\sigma}_{III_2} \right) - 4 \theta_* \dot{\sigma}_{III_2} + 1 \label{eqsigmaPIII2}
\end{equation}
with respect to the dynamics generated by \eqref{qpIII2}. By defining $\sigma_{III_2}(t) = t\frac{d}{dt} \ln \tau_{III_2}(t)$, we find the following expansions for $\tau_{III_2}(t)$: \\

\noindent \textbf{$\tau$-PIII$_{2}$ expansion} \\

\noindent The following asymptotic expansion is valid along the two rays arg$\,t = \pm \frac{\pi}{2}$ (i.e. $s \in i \mathbb{R}$) and has the form
\begin{equation}
\begin{split}
& \tau_{III_2}(t) = s^{\theta_*^2}\sum_{n \in \mathbb{Z}} e^{i n \rho} \mathcal{G}(\nu + \sqrt{3} n, s), \;\;\;\;\;\; 54 t = s^3, \\
& \mathcal{G}(\nu, s) = C(\nu,s) \left[ 1 + \sum_{k=1}^{\infty} \dfrac{D_k(\nu)}{s^k} \right], \\
& C(\nu,s) = (2\pi)^{-\frac{\nu}{2\sqrt{3}}} e^{-\frac{s^2}{8} + \nu s + \frac{i \pi \nu^2}{2} - \theta_* s} s^{\frac{1}{12} - \frac{\nu^2}{6}} 2^{-\frac{\nu^2}{3}} 3^{-\frac{\nu^2}{4}} G\left(1+\dfrac{\nu}{\sqrt{3}}\right),
\label{III21}
\end{split}
\end{equation} 
where the first few coefficients are given by
\begin{equation}
\begin{split}
D_1(\nu) = & - \dfrac{5 \nu^3}{108} + \dfrac{2 \theta_*}{3} \nu^2 - \left( 2\theta_*^2 - \dfrac{13}{72} \right) \nu + \dfrac{\theta_*(4\theta_*^2 - 1)}{3}, \\
D_2(\nu) = & \dfrac{25 \nu^6}{23328} - \dfrac{5 \theta_*}{162} \nu^5 + \dfrac{612 \theta_*^2 - 145}{1944} \nu^4 - \dfrac{\theta_*(113 \theta_*^2 - 68)}{81} \nu^3 \\
& + \left( \dfrac{26 \theta_*^4}{9} - \dfrac{35 \theta_*^2}{12} + \dfrac{767}{3456} \right) \nu^2 -
\dfrac{\theta_*(576 \theta_*^4 - 772 \theta_*^2 + 145)}{216} \nu + \dfrac{2\theta_*^2(4\theta_*^4 - 5 \theta_*^2 + 1)}{9}.
\end{split}
\end{equation}
From the number of Barnes $G$-functions we see that there is only one light particle in this sector. From these expressions we deduce
\begin{equation}
\ln \left[ \mathcal{G}\left(\frac{\nu}{\epsilon}, \frac{\theta_*}{\epsilon}, \frac{s}{\epsilon} \right) \right] \;=\;
\sum_{g \geqslant 0} \epsilon^{2g-2} \mathcal{F}_g(\nu, \theta_*, s),
\end{equation}
where we have, for instance,
\begin{equation}
\begin{split}
\mathcal{F}_0(\nu, \theta_*, s) & = -\dfrac{s^2}{8} + \nu s - \theta_* s + \dfrac{\nu^2}{6} \ln \dfrac{e^{3\pi i} \nu}{36 s} - \dfrac{\nu^2}{4} - \dfrac{5 \nu^3}{108 s} + \dfrac{2\theta_* \nu^2}{3s} - \dfrac{2\theta_*^2 \nu}{s} 
+ \dfrac{4\theta_*^3}{3s} \\
& - \dfrac{515 \nu^4}{7776 s^2} + \dfrac{19\theta_* \nu^3}{27 s^2} 
- \dfrac{7\theta_*^2 \nu^2}{3 s^2} + \dfrac{8 \theta_*^3 \nu}{3 s^2} - \dfrac{2\theta_*^4}{3 s^2} + O(s^{-3}), \\
\mathcal{F}_1(\nu, \theta_*, s) & = \zeta'(-1) - \dfrac{1}{12}\ln \dfrac{\nu}{\sqrt{3}s} + \dfrac{13 \nu}{72 s} 
- \dfrac{\theta_*}{3s} + \dfrac{533 \nu^2}{2592 s^2} - \dfrac{11\theta_* \nu}{18 s^2} + \dfrac{\theta_*^2}{6s^2} + O(s^{-3}),  \\
\mathcal{F}_2(\nu, \theta_*, s) & = - \dfrac{1}{80 \nu^2} + O(s^{-3}), \\
\mathcal{F}_g(\nu, \theta_*, s) & = \ldots.
\label{PIII2logexp1}
\end{split}
\end{equation}


\subsection{Painlev\'{e} III$_{1}$} \label{SecPIII1}

The Lax pair for PIII$_{1}$ (i.e. generic Painlev\'{e} III equation) can be obtained from the one given in \cite{JM2} by modifying it according to the discussion in \cite{Ok4}\footnote{We will not discuss the Lax pair for PV$_{deg}$; this can be found for example in \cite{2009arXiv0902.1702V}.}; explicitly, we have
\begin{equation}
\begin{split}
& {\bf A} = A_0 + \dfrac{A_1}{z}  + \dfrac{A_2}{z^2}  = \\
& = \left(\begin{array}{cc}
\frac{\sqrt{t}}{2} - \frac{\theta_*}{z} + \frac{\sqrt{t}(2p-1)}{2z^2} & \;\;\;
-\frac{p q u}{z} - \frac{\sqrt{t} p u}{z^2} \\ 
\;\; \frac{2pq - 2p^2q + 2(\theta_* + \theta_{\star}) - 4p \theta_*}{2p u z} + \frac{\sqrt{t}(p-1)}{u z^2} &
- \frac{\sqrt{t}}{2} + \frac{\theta_*}{z} - \frac{\sqrt{t}(2p-1)}{2z^2}
\end{array} \right),
\end{split} 
\end{equation}
\begin{equation}
\begin{split}
& {\bf B} = \dfrac{B_0}{z}  + B_1 + z B_2 =  \left(\begin{array}{cc}
\frac{z}{4\sqrt{t}} + \frac{1 - 2p}{4\sqrt{t} z} & 
\;\; - \frac{pqu}{2t} + \frac{p u}{2\sqrt{t} z} \\ 
\frac{2pq - 2p^2q + 2(\theta_* + \theta_{\star}) - 4p \theta_*}{4tpu} + \frac{1 - p}{2\sqrt{t} u z} & \;\;
- \frac{z}{4\sqrt{t}} - \frac{1 - 2p}{4\sqrt{t} z}
\end{array} \right). 
\end{split}
\end{equation}
The compatibility condition \eqref{cc} requires
\renewcommand\arraystretch{1.8}
\begin{equation}
\left\{
\begin{array}{l}
\dot{u} = \dfrac{u}{t} \left( \theta_* - \dfrac{\theta_* + \theta_{\star}}{p} -q  \right), \\
\dot{q} = 1 - \dfrac{q^2}{t} + \dfrac{2pq^2}{t} + \dfrac{2q\theta_*}{t}, \\
\dot{p} = \dfrac{2pq}{t} - \dfrac{2p^2q}{t} + \dfrac{\theta_* + \theta_{\star}}{t} - \dfrac{2p\theta_*}{t} ,
\end{array} \right. \label{qpIII1}
\end{equation}
\renewcommand\arraystretch{1}
and leads to the PIII$_{1}$ equation
\begin{equation}
\ddot{q} = \dfrac{\dot{q}^2}{q} - \dfrac{\dot{q}}{t} + \dfrac{q^3}{t^2} + \dfrac{2 q^2 \theta_{\star}}{t^2} 
+ \dfrac{1-2\theta_*}{t} - \dfrac{1}{q}. \label{PIII1}
\end{equation}
The trace 
\begin{equation}
\dfrac{1}{2} \operatorname{Tr}{\bf A}^2 = \dfrac{t}{4z^4} - \dfrac{\sqrt{t} \theta_{\star}}{z^3} + \dfrac{\sigma_{III_1}}{z^2} 
- \dfrac{\sqrt{t} \theta_*}{z} + \dfrac{t}{4} \label{trA2PIII1}
\end{equation}
involves the function
\begin{equation}
\sigma_{III_1}(t) = p^2q^2 - pq^2 + pt + 2pq\theta_*  - q (\theta_* + \theta_{\star}) - \frac{t}{2} + \theta_*^2.
\end{equation}
This function satisfies the $\sigma$-form of PIII$_{1}$  equation
\begin{equation}
(t\ddot{\sigma}_{III_1})^2 = (4\dot{\sigma}_{III_1}^2 - 1) \left( \sigma_{III_1} - t \dot{\sigma}_{III_1} \right) - 4 \theta_* \theta_{\star} \dot{\sigma}_{III_1} + \theta_*^2 + \theta_{\star}^2 \label{eqsigmaPIII1}
\end{equation}
with respect to the dynamics given by \eqref{qpIII1}. Introducing the tau function by $\sigma_{III_1}(t) = t\frac{d}{dt} \ln \tau_{III_1}(t)$, we find the following expansions: \\

\noindent \textbf{$\tau$-PIII$_{1}$ expansion} \\

\noindent The asymptotic series for $\tau_{III_1}(t)$ on the ray arg$\,t = 0$ (i.e. $s \in \mathbb{R}$) has the form
\begin{equation}
\begin{split}
& \tau_{III_1}(t) = s^{-\frac{1}{6} + \theta_*^2 + \theta_{\star}^2} \sum_{n \in \mathbb{Z}} e^{i n \rho} \mathcal{G}(\nu + n, s), \;\;\;\;\;\; t = \dfrac{s^2}{16}, \\
& \mathcal{G}(\nu, s) = C(\nu,s) \left[ 1 + \sum_{k=1}^{\infty} \dfrac{D_k(\nu)}{s^k} \right], \\
& C(\nu,s) = (2\pi)^{-\nu} e^{\frac{s^2}{32} + i \nu s + \frac{i \pi \nu^2}{2}} s^{- \nu^2 + \frac{1}{6} + \frac{-\theta_*^2 + 2 \theta_* \theta_{\star} - \theta_{\star}^2}{4}} 2^{-\nu^2} G\left(1+\nu + \dfrac{\theta_* - \theta_{\star}}{2}\right) G\left(1+\nu - \dfrac{\theta_* - \theta_{\star}}{2}\right),
\label{III11}
\end{split}
\end{equation} 
where the first few coefficients are given by
\begin{equation}
\begin{split}
D_1(\nu) = & - i \nu \left( \nu^2 - \dfrac{9\theta_*^2 + 14 \theta_* \theta_{\star} + 9 \theta_{\star}^2 - 2}{4} \right), \\
D_2(\nu) = & - \dfrac{\nu^6}{2} + \dfrac{9\theta_*^2 + 14 \theta_* \theta_{\star} + 9 \theta_{\star}^2 - 7}{4} \nu^4 \\
& - \left( \dfrac{\left( 9\theta_*^2 + 14 \theta_* \theta_{\star} + 9 \theta_{\star}^2 \right)^2}{32} - \dfrac{15\theta_*^2 + 26 \theta_* \theta_{\star} + 15 \theta_{\star}^2}{2} + \dfrac{15}{8} \right)\nu^2 \\
& - \dfrac{(\theta_* - \theta_{\star})^2(33\theta_*^2 + 62 \theta_* \theta_{\star} + 33 \theta_{\star}^2 - 12)}{64} .
\end{split}
\end{equation}
The number of Barnes functions implies that this sector contains an $SU(2)$ doublet of light particles. We then obtain
\begin{equation}
\ln \left[ \mathcal{G}\left(\frac{\nu}{\epsilon}, \frac{\theta_*}{\epsilon}, \frac{\theta_{\star}}{\epsilon}, 
\frac{s}{\epsilon} \right) \right] \;=\;
\sum_{g \geqslant 0} \epsilon^{2g-2} \mathcal{F}_g(\nu, \theta_*, \theta_{\star}, s),
\end{equation}
where, in particular,
\begin{equation}
\begin{split}
\mathcal{F}_0(\nu, \theta_*, \theta_{\star}, s) & = \dfrac{s^2}{32} + i\nu s 
+ \dfrac{(\nu + \theta_{*}/2 - \theta_{\star}/2)^2}{2} \ln \dfrac{\nu + \theta_{*}/2 - \theta_{\star}/2}{s} \\
& + \dfrac{(\nu - \theta_{*}/2 + \theta_{\star}/2)^2}{2} \ln \dfrac{\nu - \theta_{*}/2 + \theta_{\star}/2}{s} 
- \nu^2 \ln 2 + i \pi \dfrac{\nu^2}{2} \\
& - \dfrac{3\nu^2}{2} - \dfrac{3(\theta_* - \theta_{\star})^2}{8} 
- \dfrac{i \nu^3}{s} + \dfrac{i(9\theta_*^2 + 14\theta_* \theta_{\star} + 9\theta_{\star}^2)\nu}{4s} \\
& - \dfrac{5 \nu^4}{4 s^2} + \dfrac{(51\theta_*^2 + 90\theta_* \theta_{\star} + 51\theta_{\star}^2)\nu^2}{8 s^2} 
- \dfrac{(\theta_* - \theta_{\star})^2(33\theta_*^2 + 62\theta_* \theta_{\star} + 33\theta_{\star}^2)}{64 s^2} + O(s^{-3}), \\
\mathcal{F}_1(\nu, \theta_*, \theta_{\star}, s) & = 2 \zeta'(-1) - \dfrac{1}{12}\ln \dfrac{\nu + \theta_{*}/2 - \theta_{\star}/2}{s} - \dfrac{1}{12}\ln \dfrac{\nu - \theta_{*}/2 + \theta_{\star}/2}{s} \\
& - \dfrac{i \nu}{2 s} - \dfrac{7 \nu^2}{4 s^2} + \dfrac{3(\theta_* - \theta_{\star})^2}{16s^2} + O(s^{-3}), \\
\mathcal{F}_2(\nu, \theta_*, \theta_{\star}, s) & = - \dfrac{1}{240 (\nu + \theta_{*}/2 - \theta_{\star}/2)^2}
- \dfrac{1}{240 (\nu - \theta_{*}/2 + \theta_{\star}/2)^2} + O(s^{-3}), \\
\mathcal{F}_g(\nu, \theta_*, \theta_{\star}, s) & = \ldots.
\label{PIII1logexp1}
\end{split}
\end{equation}
The asymptotic series on the ray arg$\,t = \pi$ can be obtained from the above expansion using the symmetry $t \rightarrow -t$, $\theta_* \rightarrow -\theta_*$ of Painlev\'{e} III$_{1}$. Unlike in PII, PIV and PV case below, the quadratic term in the exponential is present in both expansions.

\subsection{Painlev\'{e} V} \label{SecPV}

The PV Lax pair is given by \cite{JM2}
\renewcommand\arraystretch{1.4}
\begin{equation}
\begin{split}
& {\bf A} = A_0 + \dfrac{A_1}{z}  + \dfrac{ A_2}{z-1} = \\
& = \left(\begin{array}{cc}
\frac{t}{2} + \frac{1}{z}(pq + \frac{\theta_0}{2}) - \frac{1}{z-1}(pq + \frac{\theta_0 + \theta_{\infty}}{2}) & 
-u\frac{pq + \theta_0}{z} + u q\frac{pq + (\theta_0 - \theta_1 + \theta_{\infty}) /2}{z-1} \\ 
\frac{pq}{z u} - \frac{pq + (\theta_0 + \theta_1 + \theta_{\infty}) /2}{(z-1)qu} \; & 
\; -\frac{t}{2} - \frac{1}{z}(pq + \frac{\theta_0}{2}) + \frac{1}{z-1}(pq + \frac{\theta_0 + \theta_{\infty}}{2})
\end{array} \right) ,
\end{split}
\end{equation}
\renewcommand\arraystretch{1}
\begin{equation}
{\bf B} = B_0 +  B_1z =
\left(\begin{array}{cc}
z/2 & -u \dfrac{pq + \theta_0 - q(pq + (\theta_0 - \theta_1 + \theta_{\infty}) /2)}{t} \\ 
\frac{1}{t u} (p q - p - \frac{\theta_0 + \theta_1 + \theta_{\infty}}{2q}) \; & \; -z/2
\end{array} \right) .
\end{equation}
The compatibility condition \eqref{cc} gives
\begin{equation}
\left\{
\begin{array}{l}
\dot{u} = \dfrac{u}{2tq}\left[ 2pq(1-q)^2 + \theta_0 + \theta_1 + \theta_{\infty} - 2q \theta_0 
+ q^2(\theta_0 - \theta_1 + \theta_{\infty}) \right], \\
\dot{q} = \dfrac{1}{2t} \left[-4pq(1-q)^2 -3\theta_0 - \theta_1 - \theta_{\infty} +2q(t + 2\theta_0 + \theta_{\infty}) -q^2(\theta_0 - \theta_1 + \theta_{\infty}) \right], \\
\dot{p} = \dfrac{1}{2t q^2} \left[2q^2p^2(1-4q+3q^2) + 2q^2 p (q(\theta_0 - \theta_1 + \theta_{\infty}) - 2\theta_0 - \theta_{\infty} - t) -\theta_0(\theta_0 + \theta_1 + \theta_{\infty}) \right],
\end{array} \right. \label{oiu}
\end{equation}
from which one can extract the PV equation
\begin{equation}
\begin{split}
\ddot{q} &= \dfrac{\dot{q}^2}{2q} + \dfrac{\dot{q}^2}{q-1} - \dfrac{\dot{q}}{t} + q\dfrac{1 - \theta_0 - \theta_{1}}{t} - q \dfrac{q+1}{2(q-1)} \\
& + \dfrac{(q-1)^2}{t^2}\left( q\dfrac{(\theta_0 - \theta_1 + \theta_{\infty})^2}{8} 
- \dfrac{(\theta_0 - \theta_1 - \theta_{\infty})^2}{8q} \right).
\end{split}
\end{equation}
The trace 
\begin{equation}
\dfrac{1}{2} \operatorname{Tr}{\bf A}^2 = \dfrac{t^2}{4} + \dfrac{\theta_0^2}{4z^2} + \dfrac{\theta_1^2}{4(z-1)^2} 
+ \dfrac{- U(t) -t \theta_{\infty}/4}{z} + \dfrac{U(t) - t \theta_{\infty}/4}{z-1} \label{trA2PV}
\end{equation}
contains the function
\begin{equation}
\begin{split}
U(t) &= -pqt -t\dfrac{\theta_0 + \theta_{\infty}}{2} + t \dfrac{\theta_{\infty}}{4} - \dfrac{\theta_0^2 + \theta_1^2 - \theta_{\infty}^2}{4} \\
&- \left( pq - \dfrac{1}{q}\left(pq + \frac{\theta_0 + \theta_1 + \theta_{\infty}}{2}\right) \right)
\left( pq + \theta_0 - q\left(pq + \frac{\theta_0 - \theta_1 + \theta_{\infty}}{2}\right) \right).
\end{split}
\end{equation}
The $\sigma$-PV equation is satisfied by the combination
\begin{equation}
\sigma_{V}(t) = U(t) + t \dfrac{\theta_{\infty}}{4} + \dfrac{\theta_0^2 + \theta_1^2 - \theta_{\infty}^2}{4} + \dfrac{t}{2} (\theta_0 + \theta_{\infty})
+ \dfrac{(\theta_0 + \theta_{\infty})^2 - \theta_1^2}{4}.
\end{equation}
It is explicitly written as
\begin{equation}
\begin{split}
(t \ddot{\sigma}_{V})^2 & = \left( \sigma_{V} - t \dot{\sigma}_{V} + 2 \dot{\sigma}_{V}^2 - (2\theta_0 + \theta_{\infty})\dot{\sigma}_{V} \right)^2 \\ 
& - 4 \dot{\sigma}_{V} \left(\dot{\sigma}_{V} - \theta_0 \right) 
\left(\dot{\sigma}_{V} - \frac{\theta_0 - \theta_1 + \theta_{\infty}}{2} \right) 
\left(\dot{\sigma}_{V} - \frac{\theta_0 + \theta_1 + \theta_{\infty}}{2} \right).  \label{pqw}
\end{split}
\end{equation}
From this one can easily extract the $\tau$-PV equation. We prefer to redefine 
\begin{equation}
\zeta_V(t) = \sigma_V(t) + \dfrac{(2\theta_0 + \theta_{\infty})^2}{8} + \dfrac{2\theta_0 + \theta_{\infty}}{4}t,
\end{equation}
and change the notation as $\theta_0 \rightarrow 2 \theta_t$, $\theta_1 \rightarrow 2\theta_0$, $\theta_{\infty} \rightarrow 2\theta_*$, so that \eqref{pqw} becomes
\begin{equation}
\left( t \ddot{\zeta}_V \right)^2 = \left( \zeta_V - t \dot{\zeta}_V + 2 \dot{\zeta}_V^2 \right)^2 
- \dfrac{1}{4}\left( \left( 2 \dot{\zeta}_V - \theta_* \right)^2 - 4 \theta_0^2 \right) \left( \left( 2 \dot{\zeta}_V + \theta_* \right)^2 - 4 \theta_t^2 \right).
\end{equation}
The function $\zeta_V(t)$ is related to the tau function via
\begin{equation}
\zeta_V(t) = t\frac{d}{dt} \ln \left( e^{-\frac{\theta_* t}{2}} t^{-\theta_0^2 - \theta_t^2 - \frac{\theta_*^2}{2}} \tau_V(t) \right). \label{avogadro}
\end{equation}
This tau function admits the following long-distance expansions along the canonical rays $\arg t=0,\pi,\pm\frac{\pi}2$:\\

\noindent \textbf{$\tau$-PV expansion 1} \\

\noindent On the rays arg$\,t = 0, \pi$ (i.e. $s \in \mathbb{R}$) we can write
\begin{equation}
\begin{split}
& \tau_V(t) = s^{-\frac{1}{3} + 2\theta_0^2 + 2 \theta_t^2 + \theta_*^2}\sum_{n \in \mathbb{Z}} e^{i n \rho} \mathcal{G}(\nu + n, s), \;\;\;\;\;\; t = 2 s, \\
& \mathcal{G}(\nu, s) = C(\nu,s) \left[ 1 + \sum_{k=1}^{\infty} \dfrac{D_k(\nu)}{s^k} \right], \\
& C(\nu,s) = (2\pi)^{-\frac{\nu}{2}} e^{\frac{s^2}{8} + i \nu s - \frac{i \pi \nu^2}{4} + \theta_* s} s^{- \frac{\nu^2}{2} + \frac{1}{12}} 2^{-\nu^2} G\left(1+\nu \right),
\end{split}
\end{equation} 
where the first few coefficients are given by
\begin{equation}
\begin{split}
D_1(\nu) = & - \dfrac{i \nu^3}{4} + \dfrac{i \nu \left( 32 \theta_0^2 + 32 \theta_t^2 + 16 \theta_*^2 - 7 \right)}{8} - 4 \theta_* \left( \theta_0^2 - \theta_t^2 \right), \\
D_2(\nu) = & - \dfrac{\nu^6}{32} + \dfrac{\left(8\theta_0^2 + 8 \theta_t^2 + 4 \theta_*^2 - 3\right) \nu^4}{8} + i \theta_* \left( \theta_0^2 - \theta_t^2 \right) \nu^3 \\
& - \left( 2 \left( 2\theta_0^2 + 2 \theta_t^2 + \theta_*^2 \right)^2 - \dfrac{19}{4} \left( 2\theta_0^2 + 2 \theta_t^2 + \theta_*^2 \right) + \dfrac{229}{128} \right)\nu^2 \\
& - \dfrac{i \theta_*(\theta_0^2 - \theta_t^2)(32\theta_0^2 + 32 \theta_t^2 + 16 \theta_*^2 - 39)\nu}{2} 
+ \left( 4\theta_*^2 - 1 \right) \left( 2 \left( \theta_0^2 - \theta_t^2 \right)^2 - \theta_0^2 - \theta_t^2 + \dfrac{1}{8} \right).
\end{split}
\end{equation}
It can be deduced from the number of Barnes functions  that there is a single light particle in this sector.
From these expressions we get
\begin{equation}
\ln \left[ \mathcal{G}\left(\frac{\nu}{\epsilon}, \frac{\theta_*}{\epsilon}, \frac{\theta_0}{\epsilon}, \frac{\theta_t}{\epsilon}, \frac{s}{\epsilon} \right) \right] \;=\;
\sum_{g \geqslant 0} \epsilon^{2g-2} \mathcal{F}_g(\nu, \theta_*, \theta_0, \theta_t, s),
\end{equation}
where 
\begin{equation}
\begin{split}
\mathcal{F}_0(\nu, \theta_*, \theta_0, \theta_t, s) & = 
\dfrac{s^2}{8} + i\nu s + \theta_* s + \dfrac{\nu^2}{2} \ln \dfrac{\nu}{4 i s} - \dfrac{3\nu^2}{4} \\
& - \dfrac{i \nu^3}{4 s} + \dfrac{2i \nu (2 \theta_0^2 + 2 \theta_t^2 + \theta_*^2)}{s} 
- \dfrac{4\theta_* (\theta_0^2 - \theta_t^2)}{s} \\
& - \dfrac{5 \nu^4}{32 s^2} + \dfrac{3(2\theta_0^2 + 2\theta_t^2 + \theta_*^2) \nu^2}{s^2} + \dfrac{16 i \theta_* (\theta_0^2 - \theta_t^2)\nu}{s^2} \\
&- \dfrac{2(\theta_0^2 - \theta_t^2)^2 + 4(\theta_0^2 + \theta_t^2)\theta_*^2}{s^2} + O(s^{-3}), \\
\mathcal{F}_1(\nu, \theta_*, \theta_0, \theta_t, s) & = 
\zeta'(-1) - \dfrac{1}{12}\ln \dfrac{\nu}{s} - \dfrac{7 i \nu}{8 s} 
- \dfrac{45 \nu^2}{32 s^2} + \dfrac{2\theta_0^2 + 2\theta_t^2 + \theta_*^2}{2s^2} + O(s^{-3}), \\
\mathcal{F}_2(\nu, \theta_*, \theta_0, \theta_t, s) & = 
- \dfrac{1}{240 \nu^2} - \dfrac{1}{8s^2} + O(s^{-3}), \\
\mathcal{F}_g(\nu, \theta_*, \theta_0, \theta_t, s) & = \ldots.
\label{PVlogexp1}
\end{split}
\end{equation} 

\noindent \textbf{$\tau$-PV expansion 2} \\

\noindent 
On the complementary rays arg$\,t = \pm \frac{\pi}{2}$ (i.e. $s \in i \mathbb{R}$) we can write
\begin{equation}
\begin{split}
& \tau_V(t) = s^{-\frac{1}{3} + 2\theta_0^2 + 2 \theta_t^2 + \theta_*^2}\sum_{n \in \mathbb{Z}} e^{i n \rho} \mathcal{G}(\nu + n, s), \;\;\;\;\;\; t = s, \\
& \mathcal{G}(\nu, s) = C(\nu,s) \left[ 1 + \sum_{k=1}^{\infty} \dfrac{D_k(\nu)}{s^k} \right], \\
& C(\nu,s) = (2\pi)^{-2\nu} e^{\nu s + \frac{\theta_* s}{2}} s^{- 2 \nu^2 + \frac{1}{3} - \theta_0^2 - \theta_t^2 - \frac{\theta_*^2}{2}} 2^{-2 \nu^2} G\left(1+\nu + \theta_0 - \dfrac{\theta_*}{2} \right)\times  \\
& \hspace{1.3 cm}\times G\left(1+\nu + \theta_t + \dfrac{\theta_*}{2} \right) G\left(1+\nu - \theta_0 - \dfrac{\theta_*}{2} \right) G\left(1+\nu - \theta_t + \dfrac{\theta_*}{2} \right),
\end{split}
\end{equation} 
where the first few coefficients are given by
\begin{equation}
\begin{split}
D_1(\nu) = & 4\nu^3 - \left( 2 \theta_0^2 + 2 \theta_t^2 + \theta_*^2 \right) \nu + \theta_* \left( \theta_t^2 - \theta_0^2 \right), \\
D_2(\nu) = & 8 \nu^6 - 2 \left(4\theta_0^2 + 4 \theta_t^2 + 2 \theta_*^2 - 5 \right) \nu^4 + 4 \theta_* \left( \theta_t^2 - \theta_0^2 \right) \nu^3 \\
& + \dfrac{1}{2} \left( 2\theta_0^2 + 2 \theta_t^2 + \theta_*^2 \right) \left( 2\theta_0^2 + 2 \theta_t^2 + \theta_*^2 - 6 \right) \nu^2 - \theta_*(\theta_t^2 - \theta_0^2)(2\theta_0^2 + 2 \theta_t^2 + \theta_*^2 - 4)\nu \\ 
& + \dfrac{4\theta_*^2 \left( \theta_t^2 - \theta_0^2 \right)^2 + \left( \theta_*^2 - 4 \theta_0^2 \right) \left( \theta_*^2 - 4 \theta_t^2 \right)}{8}.
\end{split}
\end{equation}
From the number of Barnes $G$-functions it follows that there is an $SU(4)$ quartet of light particles in this sector.
This expansion is equivalent to the one proposed in \cite[Conjecture 4.1]{Nagoya:2015}. The function $\mathcal G(\nu,s)$ is interpreted there as $c=1$ Virasoro conformal block that involves irregular vertex operators intertwining two Whittaker modules of rank 1. 
From this we recover
\begin{equation}
\ln \left[ \mathcal{G}\left(\frac{\nu}{\epsilon}, \frac{\theta_*}{\epsilon}, \frac{\theta_0}{\epsilon}, \frac{\theta_t}{\epsilon}, \frac{s}{\epsilon} \right) \right] \;=\;
\sum_{g \geqslant 0} \epsilon^{2g-2} \mathcal{F}_g(\nu, \theta_*, \theta_0, \theta_t, s),
\end{equation}
where 
\begin{equation}
\begin{split}
\mathcal{F}_0(\nu, \theta_*, \theta_0, \theta_t, s) & = 
\nu s + \frac{\theta_* s}{2} - 2\nu^2 \ln 2 
+ \dfrac{(\nu + \theta_0 - \theta_*/2)^2}{2} \ln \dfrac{\nu + \theta_0 - \theta_*/2}{s} \\
& + \dfrac{(\nu - \theta_0 - \theta_*/2)^2}{2} \ln \dfrac{\nu - \theta_0 - \theta_*/2}{s} 
+ \dfrac{(\nu + \theta_t + \theta_*/2)^2}{2} \ln \dfrac{\nu + \theta_t + \theta_*/2}{s} \\
& + \dfrac{(\nu - \theta_t + \theta_*/2)^2}{2} \ln \dfrac{\nu - \theta_t + \theta_*/2}{s} 
- \dfrac{3(4\nu^2 + 2\theta_0^2 + 2\theta_t^2 + \theta_*^2)}{4} \\
& + \dfrac{4\nu^3}{s} - \dfrac{\nu (2 \theta_0^2 + 2 \theta_t^2 + \theta_*^2)}{s} 
- \dfrac{\theta_* (\theta_0^2 - \theta_t^2)}{s} \\
& + \dfrac{10 \nu^4}{s^2}  
- \dfrac{3(2\theta_0^2 + 2\theta_t^2 + \theta_*^2) \nu^2}{s^2} - \dfrac{4 \theta_* (\theta_0^2 - \theta_t^2)\nu}{s^2} + \dfrac{(\theta_*^2 - 4 \theta_0^2)(\theta_*^2 - 4 \theta_t^2)}{8 s^2} + O(s^{-3}), \\
\mathcal{F}_1(\nu, \theta_*, \theta_0, \theta_t, s) & = 
4\zeta'(-1) - \dfrac{1}{12}\ln \dfrac{\nu + \theta_0 - \theta_*/2}{s} 
- \dfrac{1}{12}\ln \dfrac{\nu - \theta_0 - \theta_*/2}{s} \\
& - \dfrac{1}{12}\ln \dfrac{\nu + \theta_t + \theta_*/2}{s} 
- \dfrac{1}{12}\ln \dfrac{\nu - \theta_t + \theta_*/2}{s} + O(s^{-3}), \\
\mathcal{F}_2(\nu, \theta_*, \theta_0, \theta_t, s) & = 
- \dfrac{1}{240 (\nu + \theta_0 - \theta_*/2)^2} 
- \dfrac{1}{240 (\nu - \theta_0 - \theta_*/2)^2} \\
& - \dfrac{1}{240 (\nu + \theta_t + \theta_*/2)^2} 
- \dfrac{1}{240 (\nu - \theta_t + \theta_*/2)^2} + O(s^{-3}), \\
\mathcal{F}_g(\nu, \theta_*, \theta_0, \theta_t, s) & = \ldots.
\label{PVlogexp2}
\end{split}
\end{equation}

\section{Appendix B: Strong-coupling expansions for SQCD} 
\label{AppB}

In this Appendix we compute the lowest genera prepotentials $\mathcal{F}_g$ for 4d $\mathcal{N} = 2$ SQCD which were not considered in Section \ref{gauge}.

\subsection{$\mathcal{N}=2$ $SU(2)$ SQCD - $N_f = 0$}
\label{Sec30}

The Seiberg-Witten curve for $\mathcal{N} = 2$ $SU(2)$ with $N_f = 0$ reads \cite{2009arXiv0907.3987G}
\begin{equation}
y^2 = \dfrac{\Lambda^2}{z^3} + \dfrac{2u}{z^2} + \dfrac{\Lambda^2}{z}.
\end{equation}
In this representation it coincides with \eqref{trA2PIII3}. For computations it is actually more convenient to use the equivalent representation \cite{1997IJMPA..12.3413M}
\begin{equation}
y^2 = \left( x^2 - u \right)^2 - \Lambda^4.
\end{equation}
The zeroes of the discriminant
\begin{equation}
\Delta = 256 \Lambda^8 (u^2 - \Lambda^4)
\end{equation}
tell us the position of the singularities (apart the one at $u = \infty$) of the Coulomb branch moduli space; in this case, these are located at
\begin{equation}
u_1 = \Lambda^2 ,\;\;\; u_2 = -\Lambda^2 .
\end{equation}
We will therefore have two expansions for the genus zero prepotential $\mathcal{F}_0$, obtained from \eqref{F0} evaluated around $u_1$ and $u_2$ respectively; these will be related by the $\mathbb{Z}_2$ symmetry $\Lambda \rightarrow i \Lambda$ that interchanges $u_1$ with $u_2$. \\

\noindent \textbf{$N_f = 0$ expansion 1} \\

\noindent The genus zero prepotential around $u_1$ (with integration constants) can be easily computed and it is given by 
\begin{equation}
\begin{split}
2\pi i \mathcal{F}_0 \;=\; & \dfrac{a_D^2}{4} \ln \left( 2^8 \dfrac{(\Lambda e^{-i\pi/2})^2}{a_D^2} \right)
+ \dfrac{3}{4} a_D^2 + b_1 a_D \Lambda + b_2 \Lambda^2 \\
& + \dfrac{a_D^3}{16 (\Lambda e^{-i\pi/2})} - \dfrac{5a_D^4}{512(\Lambda e^{-i\pi/2})^2} + \dfrac{11 a_D^5}{4096 (\Lambda e^{-i\pi/2})^3} + \ldots
\end{split}
\end{equation}
The genus one contribution is
\begin{equation}
\begin{split}
\mathcal{F}_1 \;=\; & 
- \dfrac{1}{12} \ln \dfrac{ a_D}{4\Lambda e^{-i\pi/2}}
+ \dfrac{a_D}{2^5 (\Lambda e^{-i\pi/2})} - \dfrac{3 a_D^2}{2^9 (\Lambda e^{-i\pi/2})^2} + O(\Lambda^{-3}).
\end{split}
\end{equation}
These agree with the first expansion of Section \ref{SecPIII3} for $\nu = i a_D$ and $s = 4i \Lambda$. Notice also that the Painlev\'{e} asymptotics determines the constants $b_1$ and $b_2$ appearing in the gauge theory computation of $\mathcal{F}_0$. \\
 
\noindent \textbf{$N_f = 0$ expansion 2} \\

\noindent Similarly, the genus zero prepotential around $u_2$ (again with integration constants) reads
\begin{equation}
\begin{split}
2\pi i \mathcal{F}_0 \;=\; & \dfrac{a_D^2}{4} \ln \left( 2^8 \dfrac{(\Lambda e^{i \pi/2})^2}{a_D^2} \right)
+ \dfrac{3}{4} a_D^2 + + b_1' a_D \Lambda + b_2' \Lambda^2 \\
& + \dfrac{a_D^3}{16 (\Lambda e^{i \pi/2})} - \dfrac{5a_D^4}{512(\Lambda e^{i \pi/2})^2} + \dfrac{11 a_D^5}{4096 (\Lambda e^{i \pi/2})^3} + \ldots
\end{split}
\end{equation}
The genus one
\begin{equation}
\begin{split}
\mathcal{F}_1 \;=\; 
- \dfrac{1}{12} \ln \dfrac{ a_D}{4\Lambda e^{i\pi/2}} + \dfrac{a_D}{2^5 (\Lambda e^{i\pi/2})} - \dfrac{3 a_D^2}{2^9 (\Lambda e^{i\pi/2})^2} + O(\Lambda^{-3}).
\end{split}
\end{equation}
These are the same expressions as above with $\frac{a_D}{\Lambda} \rightarrow - \frac{a_D}{\Lambda}$.


\subsection{$\mathcal{N}=2$ $SU(2)$ SQCD - $N_f = 1$}
\label{Sec31}

The Seiberg-Witten curve for $\mathcal{N} = 2$ $SU(2)$ with $N_f = 1$ is given by \cite{2009arXiv0907.3987G}
\begin{equation}
y^2 = \dfrac{\Lambda^2}{z^3} + \dfrac{3u}{z^2} + \dfrac{2\Lambda m}{z} + \Lambda^2.
\end{equation}
In this representation it coincides with \eqref{trA2PIII2}. For computations we will use the equivalent representation \cite{1997IJMPA..12.3413M}
\begin{equation}
y^2 = \left( x^2 - u \right)^2 - \Lambda^3 (x + m).
\end{equation}
The zeroes of the discriminant
\begin{equation}
\Delta = -\Lambda^6 (256u^3 - 256 m^2 u^2 - 288 m u \Lambda^3 + 256 m^3 \Lambda^3 + 27 \Lambda^3)
\end{equation}
are located at (perturbatively in $m$ small)
\begin{equation}
\begin{split}
u_1 &= -\dfrac{3}{4}\dfrac{\Lambda^2}{2^{2/3}} - \dfrac{m\Lambda}{2^{1/3}} + \dfrac{m^2}{3} + O(m^3), \\
u_2 &= -\dfrac{3}{4}\dfrac{(e^{2\pi i/3}\Lambda)^2}{2^{2/3}} - \dfrac{m(e^{2\pi i/3}\Lambda)}{2^{1/3}} + \dfrac{m^2}{3} + O(m^3), \\
u_3 &= -\dfrac{3}{4}\dfrac{(e^{4\pi i/3}\Lambda)^2}{2^{2/3}} - \dfrac{m(e^{4\pi i/3}\Lambda)}{2^{1/3}} + \dfrac{m^2}{3} + O(m^3).
\end{split}
\end{equation}
We therefore expect three expansions related by $\mathbb{Z}_3$ symmetry. \\

\noindent \textbf{$N_f = 1$ expansion 1} \\

\noindent The genus zero prepotential around $u_1$ reads
\begin{equation}
\begin{split}
2\pi i \mathcal{F}_0 = & \dfrac{a_D^2}{4} \ln \left( 2^{7/3}3^5 \dfrac{(\Lambda e^{-i \pi/2})^2}{a_D^2} \right)
+ \dfrac{3}{4} a_D^2 + b_1(\Lambda, m) a_D + b_2(\Lambda, m) \\
& - \dfrac{5 a_D^3}{18 \sqrt{3}\, 2^{1/6} (\Lambda e^{-i \pi/2})} + \dfrac{2^{4/3} (im) a_D^2}{3 (\Lambda e^{-i \pi/2})} \\
& + \dfrac{515a_D^4}{2^{1/3} 1944(\Lambda e^{-i \pi/2})^2} - \dfrac{2^{1/6} 38 (im) a_D^3}{27 \sqrt{3} (\Lambda e^{-i \pi/2})^2} + \dfrac{2^{2/3} 7 (im)^2 a_D^2}{9 (\Lambda e^{-i \pi/2})^2} \\
& - \dfrac{10759 a_D^5}{17496 \sqrt{6} (\Lambda e^{-i \pi/2})^3} + \dfrac{805 (im) a_D^4}{729 (\Lambda e^{-i \pi/2})^3} - \dfrac{55\sqrt{2} (im)^2 a_D^3}{27 \sqrt{3} (\Lambda e^{-i \pi/2})^3} + \dfrac{80 (im)^3 a_D^2}{81 (\Lambda e^{-i \pi/2})^3} + \ldots.
\end{split}
\end{equation}
The genus one contribution is
\begin{equation}
\begin{split}
\mathcal{F}_1 \;=\; & 
- \dfrac{1}{12} \ln \dfrac{2 a_D}{(2^{1/6} 3 \Lambda e^{-i\pi/2})} 
- \dfrac{13 a_D}{2^{1/6}36 \sqrt{3} (\Lambda e^{-i\pi/2})} + \dfrac{2^{1/3}(i m)}{9(\Lambda e^{-i\pi/2})} \\
& + \dfrac{533 a_D^2}{2^{1/3}1944 (\Lambda e^{-i\pi/2})^2} - \dfrac{2^{1/6}11(i m)a_D}{27 \sqrt{3}(\Lambda e^{-i\pi/2})^2}
+ \dfrac{(i m)^2}{2^{1/3}27(\Lambda e^{-i\pi/2})^2} + O(\Lambda^{-3}).
\end{split}
\end{equation}

\noindent \textbf{$N_f = 1$ expansion 2} \\

\noindent The genus zero prepotential around $u_2$ reads
\begin{equation}
\begin{split}
2\pi i \mathcal{F}_0 = & \dfrac{a_D^2}{4} \ln \left( 2^{7/3}3^5 \dfrac{(\Lambda e^{2 i \pi/3})^2}{a_D^2} \right)
+ \dfrac{3}{4} a_D^2 + b_1'(\Lambda, m) a_D + b_2'(\Lambda, m) \\
& - \dfrac{5 a_D^3}{18 \sqrt{3}\, 2^{1/6} (\Lambda e^{2 i \pi/3})} + \dfrac{2^{4/3} (-m) a_D^2}{3 (\Lambda e^{2i \pi/3})} \\
& + \dfrac{515a_D^4}{2^{1/3} 1944(\Lambda e^{2i \pi/3})^2} - \dfrac{2^{1/6} 38 (-m) a_D^3}{27 \sqrt{3} (\Lambda e^{2i \pi/3})^2} + \dfrac{2^{2/3} 7 (-m)^2 a_D^2}{9 (\Lambda e^{2i \pi/3})^2} \\
& - \dfrac{10759 a_D^5}{17496 \sqrt{6} (\Lambda e^{2i \pi/3})^3} + \dfrac{805 (-m) a_D^4}{729 (\Lambda e^{2i \pi/3})^3} - \dfrac{55\sqrt{2} (-m)^2 a_D^3}{27 \sqrt{3} (\Lambda e^{2i \pi/3})^3} + \dfrac{80 (-m)^3 a_D^2}{81 (\Lambda e^{2i \pi/3})^3} + \ldots,
\end{split}
\end{equation}
while the genus one is
\begin{equation}
\begin{split}
\mathcal{F}_1 \;=\; & 
- \dfrac{1}{12} \ln \dfrac{2 a_D}{(2^{1/6} 3 \Lambda e^{2i\pi/3})}  
- \dfrac{13 a_D}{2^{1/6}36 \sqrt{3} (\Lambda e^{2 i \pi/3})} + \dfrac{2^{1/3}(- m)}{9(\Lambda e^{2 i \pi/3})} \\
& + \dfrac{533 a_D^2}{2^{1/3}1944 (\Lambda e^{2 i \pi/3})^2} - \dfrac{2^{1/6}11(-m)a_D}{27 \sqrt{3}(\Lambda e^{2 i \pi/3})^2}
+ \dfrac{(- m)^2}{2^{1/3}27(\Lambda e^{2 i \pi/3})^2} + O(\Lambda^{-3}).
\end{split}
\end{equation}

\noindent \textbf{$N_f = 1$ expansion 3} \\

\noindent The genus zero prepotential around $u_3$ reads
\begin{equation}
\begin{split}
2\pi i \mathcal{F}_0 = & \dfrac{a_D^2}{4} \ln \left( 2^{7/3}3^5 \dfrac{(\Lambda e^{i \pi/3})^2}{a_D^2} \right) 
+ \dfrac{3}{4} a_D^2 + b_1''(\Lambda, m) a_D + b_2''(\Lambda, m) \\
& - \dfrac{5 a_D^3}{18 \sqrt{3}\, 2^{1/6} (\Lambda e^{i \pi/3})} + \dfrac{2^{4/3} m a_D^2}{3 (\Lambda e^{i \pi/3})} \\
& + \dfrac{515a_D^4}{2^{1/3} 1944(\Lambda e^{i \pi/3})^2} - \dfrac{2^{1/6} 38 m a_D^3}{27 \sqrt{3} (\Lambda e^{i \pi/3})^2} + \dfrac{2^{2/3} 7 m^2 a_D^2}{9 (\Lambda e^{i \pi/3})^2} \\
& - \dfrac{10759 a_D^5}{17496 \sqrt{6} (\Lambda e^{i \pi/3})^3} + \dfrac{805 m a_D^4}{729 (\Lambda e^{i \pi/3})^3} - \dfrac{55\sqrt{2} m^2 a_D^3}{27 \sqrt{3} (\Lambda e^{i \pi/3})^3} + \dfrac{80 m^3 a_D^2}{81 (\Lambda e^{i \pi/3})^3} + \ldots,
\end{split}
\end{equation}
while the genus one is
\begin{equation}
\begin{split}
\mathcal{F}_1 \;=\; & 
- \dfrac{1}{12} \ln \dfrac{2 a_D}{(2^{1/6} 3 \Lambda e^{i\pi/3})} 
- \dfrac{13 a_D}{2^{1/6}36 \sqrt{3} (\Lambda e^{i \pi/3})} + \dfrac{2^{1/3}m}{9(\Lambda e^{i \pi/3})} \\
& + \dfrac{533 a_D^2}{2^{1/3}1944 (\Lambda e^{i \pi/3})^2} - \dfrac{2^{1/6}11 m a_D}{27 \sqrt{3}(\Lambda e^{i \pi/3})^2}
+ \dfrac{m^2}{2^{1/3}27(\Lambda e^{i \pi/3})^2} + O(\Lambda^{-3}).
\end{split}
\end{equation}
These three expansions agree with the results of Section \ref{SecPIII2} under identification $\nu = \mp i \sqrt{3} a_D$, $s = \pm i 2^{-5/6} 3 \Lambda e^{i \theta_{\Lambda}}$ and $\theta_* = \mp i 2^{-1/2} m e^{i \theta_m}$, with $\theta_{\Lambda} = - \frac{\pi}{2}$, $\theta_m = \frac{\pi}{2}$ for the first expansion, $\theta_{\Lambda} = \frac{2\pi}{3}$, $\theta_m = \pi$ for the second expansion and $\theta_{\Lambda} = \frac{\pi}{3}$, $\theta_m = 0$ for the third expansion. 


\subsection{$\mathcal{N}=2$ $SU(2)$ SQCD - $N_f = 2$}
\label{Sec32}

The Seiberg-Witten curve for $\mathcal{N} = 2$ $SU(2)$ with $N_f = 2$ is given by \cite{2009arXiv0907.3987G}
\begin{equation}
\begin{split}
& y^2 = \dfrac{\Lambda^2}{z^4} + \dfrac{2\Lambda m_1}{z^3} + \dfrac{4u}{z^2} + \dfrac{2\Lambda m_2}{z} + \Lambda^2 \;\;\;(\text{first realization}) \\
& y^2 = \dfrac{m^2_+}{z^2} + \dfrac{m^2_-}{(z-1)^2} + \dfrac{\Lambda^2 + u}{2z} + \dfrac{\Lambda^2 - u}{2(z-1)} \;\;\;(\text{second realization}) \\
\end{split}
\end{equation}
In the first realization it coincides with \eqref{trA2PIII1}; the second realization can be shown to coincide with the spectral curve for PV$_{deg}$ by making use of the explicit expression for its Lax pair given for example in \cite{2009arXiv0902.1702V}. For computations we will use the equivalent representation \cite{1997IJMPA..12.3413M}
\begin{equation}
y^2 = \left( x^2 - u + \dfrac{\Lambda^2}{8} \right)^2 - \Lambda^2 (x + m_1) (x + m_2).
\end{equation}
The zeroes of the discriminant
\begin{equation}
\Delta = 16 B^4 D + 256 D^3 - 128 B^2 D^2 - 4 B^3 C^2 - 27 C^4 + 144 B C^2 D
\end{equation}
with
\begin{equation}
\begin{split}
B & = -2u - \dfrac{3\Lambda^2}{4}, \\
C & = -\Lambda^2(m_1 + m_2), \\
D & = u^2 - \dfrac{u \Lambda^2}{4} + \dfrac{\Lambda^4}{64} - \Lambda^2 m_1 m_2,
\end{split}
\end{equation}
are located at (perturbatively in $m_1$, $m_2$ small)
\begin{equation}
\begin{split}
u_1 & = - \frac{\Lambda^2}{8} - \dfrac{\Lambda(m_1 + m_2)}{2} + \dfrac{(m_1 - m_2)^2}{4} + O(\Lambda^{-1}), \\
u_2 & = - \frac{\Lambda^2}{8} + \dfrac{\Lambda(m_1 + m_2)}{2} + \dfrac{(m_1 - m_2)^2}{4} + O(\Lambda^{-1}), \\
u_3 & = \frac{\Lambda^2}{8} + \dfrac{i\Lambda(m_1 - m_2)}{2} + \dfrac{(m_1 + m_2)^2}{4} + O(\Lambda^{-1}), \\
u_4 & = \frac{\Lambda^2}{8} - \dfrac{i\Lambda(m_1 - m_2)}{2} + \dfrac{(m_1 + m_2)^2}{4} + O(\Lambda^{-1}). \\
\end{split}
\end{equation}
We therefore expect four expansions for generic values of the masses. Unfortunately, computations with all masses turned on are quite cumbersome; here we present the results for $\mathcal{F}_0$ and $\mathcal{F}_1$ at zero masses (in which $u_1 = u_2$ and $u_3 = u_4$), and later we will give the expression for $\mathcal{F}_1$ at generic masses. \\

\noindent \textbf{$N_f = 2$ expansion 1,2 (massless)} \\

\noindent The genus zero prepotential around $u_1 = u_2$ (with integration constants) can be easily computed and it is given by 
\begin{equation}
\begin{split}
2\pi i \mathcal{F}_0 \;=\; & \dfrac{a_D^2}{2} \ln \left( - \dfrac{8 \Lambda^2}{a_D^2} \right)
+ \dfrac{3}{2} a_D^2 + b_1 a_D \Lambda + b_2 \Lambda^2  + \dfrac{i a_D^3}{\sqrt{2}\Lambda} + \dfrac{5a_D^4}{8\Lambda^2} + \ldots
\end{split}
\end{equation}
The genus one contribution is
\begin{equation}
\begin{split}
\mathcal{F}_1 \;=\; & 
- \dfrac{1}{6} \ln \dfrac{a_D}{\sqrt{2}\Lambda}
- \dfrac{i a_D}{2 \sqrt{2} \Lambda} - \dfrac{7 a_D^2}{8 \Lambda^2} + O(\Lambda^{-3}).
\end{split}
\end{equation}
These expressions coincide with the massless case of the PIII$_1$ expansion of Section \ref{SecPIII1} under iden\-tification $\nu = i a_D$, $s = i\sqrt{2}  \Lambda$. \\

\noindent \textbf{$N_f = 2$ expansion 3,4 (massless)} \\

\noindent The genus zero prepotential around $u_3 = u_4$ (with integration constants)  is given by the expansion
\begin{equation}
\begin{split}
2\pi i \mathcal{F}_0 \;=\; & \dfrac{a_D^2}{2} \ln \left( -\dfrac{8 \Lambda^2}{a_D^2} \right)
+ \dfrac{3}{2} a_D^2 + b_1' a_D \Lambda + b_2' \Lambda^2  - \dfrac{i a_D^3}{\sqrt{2}\Lambda} + \dfrac{5a_D^4}{8\Lambda^2} + \ldots
\end{split}
\end{equation}
The genus one counterpart is
\begin{equation}
\begin{split}
\mathcal{F}_1 \;=\; & 
- \dfrac{1}{6} \ln \dfrac{-a_D}{\sqrt{2}\Lambda}
+ \dfrac{i a_D}{2 \sqrt{2} \Lambda} - \dfrac{7 a_D^2}{8 \Lambda^2} + O(\Lambda^{-3}).
\end{split}
\end{equation}
These expansions match the massless case of the expansion of Section \ref{SecPIII1} for $\nu = -i a_D$, $s = i\sqrt{2} \Lambda$. \\

\noindent \textbf{$N_f = 2$ expansion 1 (massive)} \\

\noindent Genus one prepotential around $u_1$ (modulo constants in the logarithms):
\begin{equation}
\begin{split}
\mathcal{F}_1 \;=\; & - \dfrac{1}{12} \ln \left( \dfrac{\tilde{a}_D - i \frac{m_1+m_2}{2\sqrt{2}}}{\sqrt{2}\Lambda} \right) 
 - \dfrac{1}{12} \ln \left( \dfrac{\tilde{a}_D + i \frac{m_1+m_2}{2\sqrt{2}}}{\sqrt{2}\Lambda} \right) 
- \dfrac{i \tilde{a}_D}{2 \sqrt{2} \Lambda} \\
& - \dfrac{7 \tilde{a}_D^2}{8\Lambda^2} - \dfrac{3 (m_1 + m_2)^2}{64\Lambda^2} + O(\Lambda^{-3}),
\end{split}
\end{equation}
with 
\begin{equation}
\tilde{a}_D = a_D - i \dfrac{m_1+m_2}{2\sqrt{2}}.
\end{equation}
This coincides with the massive case of the expansion of Section \ref{SecPIII1} under identification $\nu = i \tilde{a}_D$, $s = \sqrt{2} i \Lambda$ and $\theta_* - \theta_{\star} = \frac{m_1 + m_2}{\sqrt{2}}$. The results for the other three expansions are very similar
and can be obtained by a change of signs in the parameters.


\subsection{$\mathcal{N}=2$ $SU(2)$ SQCD - $N_f = 3$}

The Seiberg-Witten curve for $\mathcal{N} = 2$ $SU(2)$ with $N_f = 3$ is given by \cite{2009arXiv0907.3987G}
\begin{equation}
y^2 = \dfrac{m^2_+}{z^2} + \dfrac{m^2_-}{(z-1)^2} + \dfrac{2\Lambda m + u}{2z} + \dfrac{2\Lambda m - u}{2(z-1)} + \Lambda^2.
\end{equation}
In this representation it coincides with \eqref{trA2PV}. For computations we will use the equivalent representation \cite{1997IJMPA..12.3413M}
\begin{equation}
y^2 = \left( x^2 - u + \dfrac{\Lambda}{4}\left(x + \frac{m_1 + m_2 + m_3}{2}\right) \right)^2 - \Lambda (x + m_1) (x + m_2) (x + m_3).
\end{equation}
Here we will only consider the massless case, in which the zeroes of the discriminant \eqref{Delta} 
are located at 
\begin{equation}
\begin{split}
u_1 = \dfrac{\Lambda^2}{256} \;\;\;,\;\;\;
u_2 = u_3 = u_4 = u_5 = 0.
\end{split}
\end{equation}
We will therefore have two different expansions. \\

\noindent \textbf{$N_f = 3$ expansion 1 (massless)} \\

\noindent The genus zero prepotential around $u_1$ is given by
\begin{equation}
\begin{split}
2\pi i \mathcal{F}_0 \;=\; & \dfrac{a_D^2}{4} \ln \left( - \dfrac{\Lambda^2}{8 a_D^2} \right)
+ \dfrac{3 a_D^2}{4} + b_1 a_D \Lambda + b_2 \Lambda^2  - \dfrac{i 2 \sqrt{2} a_D^3}{\Lambda} + \dfrac{20 a_D^4}{\Lambda^2} + \ldots.
\end{split}
\end{equation}
The genus one reads instead
\begin{equation}
\begin{split}
\mathcal{F}_1 \;=\; & 
- \dfrac{1}{12} \ln \left( - \dfrac{8 \sqrt{2} a_D}{\Lambda} \right) + \dfrac{i 7 \sqrt{2} a_D}{\Lambda} 
- \dfrac{180 a_D^2}{\Lambda^2} + O(\Lambda^{-3}).
\end{split}
\end{equation}
This can be matched with the results of the massless case of the first expansion in Section~\ref{SecPV} via $\nu = -i a_D$, $s = \frac{i \Lambda}{8\sqrt{2}}$. \\

\noindent \textbf{$N_f = 3$ expansion 2,3,4,5 (massless)} \\

\noindent The genus zero prepotential around $u_2 = u_3 = u_4 = u_5$ is given by
\begin{equation}
\begin{split}
2\pi i \mathcal{F}_0 \;=\; & a_D^2 \ln \left( - \dfrac{\Lambda^2}{32 a_D^2} \right)
+ 3 a_D^2 + b_1' a_D \Lambda + b_2' \Lambda^2  + \dfrac{i 16 \sqrt{2} a_D^3}{\Lambda} + \dfrac{320 a_D^4}{\Lambda^2} + \ldots,
\end{split}
\end{equation}
while the genus one reads
\begin{equation}
\begin{split}
\mathcal{F}_1 \;=\; & 
- \dfrac{1}{3} \ln \left( - \dfrac{4 \sqrt{2} i a_D}{\Lambda} \right) + O(\Lambda^{-3}).
\end{split}
\end{equation}
This can be matched with the results of the massless case of the second expansion in Section~\ref{SecPV} by identifying $\nu = i a_D $, $s = -\frac{ \Lambda}{4 \sqrt{2}}$.

\bibliography{bibl}
\bibliographystyle{JHEP}

\end{document}